\documentclass[aps,pre,showpacs,amsmath,amssymb,twocolumn,longbibliography,superscriptaddress]{revtex4-1}

\usepackage{graphicx}
\usepackage{epstopdf}
\usepackage{bm}
\usepackage{braket}
\usepackage{dsfont}
\usepackage[usenames,dvipsnames]{xcolor}
\usepackage[tight]{subfigure}
\usepackage{verbatim}
\usepackage{units}
\usepackage{natbib}
\usepackage{multirow}
\usepackage{enumitem}
\usepackage{mathrsfs}
\usepackage{leftidx}
\usepackage{xspace}

\usepackage[a4paper]{hyperref}
\hypersetup{colorlinks=true,linktoc=all,linkcolor=blue,breaklinks=true,citecolor=blue,urlcolor=blue}
\usepackage[english]{babel}
\AtBeginDocument{%
    \newwrite\bibnotes
    \def\bibnotesext{Notes.bib}
    \immediate\openout\bibnotes=\jobname\bibnotesext
    \immediate\write\bibnotes{@CONTROL{REVTEX41Control}}
    \immediate\write\bibnotes{@CONTROL{%
    apsrev41Control,author="08",editor="1",pages="1",title="0",year="1"}}
     \if@filesw
     \immediate\write\@auxout{\string\citation{apsrev41Control}}%
    \fi
}%

\usepackage{hyperref}

\newcommand{\tr}{{\rm Tr}}

\begin{document}

\title{Collective effects on the performance and stability of quantum heat engines}
\author{Leonardo da Silva Souza}
\email{leonardosilvasouza@id.uff.br}
\affiliation{Instituto de F\'{\i}sica, Universidade Federal Fluminense, 24210-346 Niter\'oi, Brazil}
\affiliation{Departamento de F\'{\i}sica - ICEx - Universidade Federal de Minas Gerais,
Av. Pres. Ant\^onio Carlos 6627 - Belo Horizonte - MG - Brazil - 31270-901.}

\author{Gonzalo Manzano}
\affiliation{Institute for Cross-Disciplinary Physics and Complex Systems (IFISC) UIB-CSIC, Campus Universitat Illes Balears, E-07122 Palma de Mallorca, Spain}
\affiliation{Institute for Quantum Optics and Quantum Information (IQOQI), Austrian Academy of Sciences, Boltzmanngasse 3, 1090 Vienna, Austria.}

\author{Rosario Fazio}
\affiliation{International Centre for Theoretical Physics ICTP, Strada Costiera 11, I-34151, Trieste, Italy}
\affiliation{Dipartimento di Fisica, Universit\`a di Napoli ``Federico II'', Monte S. Angelo, I-80126 Napoli, Italy}
 
 \author{Fernando Iemini}
\affiliation{Instituto de F\'{\i}sica, Universidade Federal Fluminense, 24210-346 Niter\'oi, Brazil}
\affiliation{International Centre for Theoretical Physics ICTP, Strada Costiera 11, I-34151, Trieste, Italy}

\begin{abstract}
 Recent predictions for quantum-mechanical enhancements in the operation of small heat engines have raised renewed interest in their study from both a fundamental perspective and in view of applications. One essential question is whether collective effects may help to carry enhancements over larger scales, when increasing the number of systems composing the working substance of the engine. Such enhancements may consider not only power and efficiency, that is its performance, but, additionally, its constancy, i.e. the stability of the engine with respect to unavoidable environmental fluctuations. We explore this issue by introducing a many-body quantum heat engine model composed by spin pairs working in continuous operation. We study how power, efficiency and constancy scale with the number of spins composing the engine and introduce a well-defined macroscopic limit where analytical expressions are obtained. Our results predict power enhancements, both in finite-size and macroscopic cases, for a broad range of system parameters and temperatures, without compromising the engine efficiency, accompanied by 
 coherence-enhanced constancy for 
 finite sizes. We discuss these quantities in connection to Thermodynamic Uncertainty Relations (TUR).
\end{abstract}

\maketitle

\section{Introduction}

The quest for an efficient managing and control of heat at the nanoscale~\cite{hanggi2009,benenti2017,pekola2019} has boosted, in the last years, theoretical and experimental investigations of quantum heat engines~\cite{anders2016,goold2016,binder2019}. In essence, a thermal machine consists of a finite system (the ``working medium") connected to two or more reservoirs that are kept at different temperatures. If the system is composed by a few-level quantum system, the characteristics of the thermal machine may carry distinct features of quantum mechanics. Since the Scovil and Schulz-DuBois pioneering proposals of a continuous heat engine based on the three-level maser~\cite{scovil1959,geusic1959} 
the field has burst. A plethora of models have been proposed and analyzed which, as in the case of their macroscopic counterparts, may operate in a continuous mode~\cite{kosloff2014,Mitchison2019} or in a many-strokes fashion~\cite{quan2007,kosloff2017}, paving the way to experimental implementations of quantum heat engines in the laboratory~\cite{Brantut2013,rossnagel2016,cottet2017,maslennikov2019,klatzow2019,peterson2019,lindenfels2019,Horne2020,bouton2021}.
In this context, continuous engines operating in steady-state conditions offer the advantage of avoiding a precise control over time of the working substance and their coupling and decoupling from the environment, which is instead often required to implement cycles with many strokes, and that, in practice, may incur in extra thermodynamic costs.

Understanding the operation principle of small (quantum) heat engines is , however, not a mere technological challenge: new, fundamental questions arise. Fluctuations, both classical and quantum, cannot any longer be disregarded~\cite{fluctuations1,fluctuations2}, but they become 
a key ingredient to characterize small thermal machines~\cite{Verley2014,campisi2015,Martinez2016,manzano2018,Friedman2018}. The quantumness of a thermal machine has been also a subject of an intense study , aimed to understand the role of quantum mechanics (coherence and entanglement) in determining, and possibly enhancing, the performance of heat engines and refrigerators~\cite{Scully2011,Park2013,brunner2014,Correa2014,uzdin2015,killoran2015,Brandner2017, Hammam2021}. In a similar spirit, the question of fundamental limits to the functionality of a thermal machine became relevant as well~\cite{linden2010,brunner2012,correa2015,hofer2016,Clivaz2019,Monsel2020}. 

In addition to the properties of the deep quantum regime, where the working medium is constituted by few interacting qubits 
(or qudits), understanding how heat engines approach the macroscopic limit may be also important.  This problem can be seen from different perspectives. The connection to the more general 
quantum/classical crossover is immediate. More specifically to the field of thermal engines, a crucial point is to understand how the performance of the machine scales with the dimensions (physical dimensions, size of the Hilbert space, etc) 
of the working medium~\cite{silva2016,PhysRevE.89.042128}. A simple framework to formulate this question is to consider how the power and efficiency 
of a heat engine changes with the number of units (e.g. spins, qudits or harmonic oscillators) that form the working medium, and compare it with an analogous engine composed by the same number of units, but where the units work in parallel, independently from each other. In the latter case, 
the power output will simply be given by the power of a single unit multiplied by the number $N$ of units, and 
no new emerging phenomenon is expected when $N$ approaches the macroscopic limit.

The situation may radically change when collective effects come into play, which may lead to the enhancement of thermodynamic properties of the engine with respect to the independent ``parallel'' case, as has been recently reported in a number of works. The scaling of the heat capacity can increase beyond linear if the working medium is on the verge of a phase transition, leading to a boost in the efficiency~\cite{campisi2016, Abiuso2020}. Cooperative effects have been found in quantum cycles whose working substance consist in a many-body system, such as interacting Bose gases~\cite{jaramillo2016,chen2019}, spin systems~\cite{cakmak2016,Ma2017,hardal2018,kloc2019,halpern2019,latune2020,kloc2021}, qutrits~\cite{gelbwaser2019} or bosonic models~\cite{watanabe2020}. Similar phenomena have been also found in a continuous many-body Floquet engine~\cite{niedenzu2018,Kamimura2021}, 
as well as in classical machines~\cite{vroylandt2017,herpich2018}. The cooperative effects are usually manifested as model-dependent enhancements in the output power that may eventually increase the efficiency of the collective engine, owing to infinite-range pairwise interactions between the working substance units, or by means of collective dissipation~\cite{manzano2019}, leading to superradiance-like behavior~\cite{dicke1954,haroche1982}. 
However, previous works did not take into account several elements that might make such enhancements spurious, namely, the existence of a well-defined macroscopic (thermodynamic) limit, the behavior of the Lindbladian gap ensuring a proper steady-state regime of operation, and the impact of 
increasing fluctuations in the energy currents, that may spoil the performance of the engine.


Beyond enhancements in power output or efficiency, a third key element determining the performance of a heat engine is given by its constancy, that is, the stability of the output power with respect to fluctuations~\cite{PhysRevLett.120.190602,Holubec_2014,PhysRevE.96.030102}. These three quantities (power, efficiency and constancy) are in general not independent from each other, but verify specific trade-off relations following from the so-called Thermodynamic Uncertainty Relation (TUR)~\cite{PhysRevLett.114.158101,PhysRevLett.116.120601,Horowitz:2020}. The TUR put strict constraints on the constancy achievable by any classical steady-state engine. In particular, it implies that reaching Carnot efficiency at finite power may only be possible at the expense of diverging fluctuations~\cite{PhysRevLett.120.190602}. Constancy-enhanced engines aim hence to operate at reduced power fluctuations by overcoming the TUR. This is e.g. the case of some models of quantum heat engines~\cite{PhysRevB.98.085425,PhysRevE.99.062141,PhysRevE.103.012133,kalaee2021violating} and work-converters~\cite{PhysRevB.102.165418} (see also Refs.~\cite{PhysRevB.98.155438,PhysRevE.100.042101,PhysRevB.101.195423} for  a discussion on transport setups). Extensions of the original TUR for quantum dynamics have been also recently reported in different regimes~\cite{PhysRevLett.121.130601,PhysRevResearch.1.033021,PhysRevLett.122.130605,PhysRevLett.123.090604,PhysRevLett.125.050601,PhysRevLett.126.010602}. In this context, the impact of collective effects in the constancy of quantum heat engines and its connection to the TUR's is an intriguing open question.

In this paper we 
 study how cooperative effects may enhance the power, efficiency and constancy of thermal machines by scaling up a two-qubit engine model that constitutes one of the simplest models of small quantum heat engines working in continuous operation~\cite{brunner2012,kosloff2014}. We consider two ensembles of $N$ spin$-1/2$ particles with different energy spacing, collectively coupled to respective common thermal baths at different temperatures, and subjected to a collective coherent drive that performs or extracts work from the system. Importantly, the model we introduce in this work can show violations of the classical TUR and it admits a cristal clear comparison to the case of $N$ separate two-qubit engines working in parallel.
 Moreover it is, at same time, experimentally relevant for small/moderate $N$ and amenable of a \textit{quasi-analytic} treatment in the large $N$ limit. We obtain that the output power can show super-linear enhancements at constant efficiency for moderate values of $N$ coming, however, at the cost of increased fluctuations that are detrimental to the engine constancy. Larger system sizes are also considered by introducing a proper renormalization of the model parameters in a high-temperature regime, for which linear enhancements of the power output are obtained, and, remarkably, coherence-enhanced constancy is observed for finite system sizes. In the macroscopic limit TUR violations dissapear, shedding new light on the quantum to classical crossover.

The manuscript is organized as follows. In Sec.\eqref{sec.models} we define the quantum heat engine studied in this work and its GKS-Lindblad equation for the dynamics. In Sec.\eqref{sec.therm.properties} we define the thermodynamic properties we focus our studies, namely, the work output, heat currents, efficiency and constancy of the heat engine. We also discuss these quantities in connection to the TUR and to the first and second laws at steady state conditions.
 In Sec.\eqref{sec.finite-size} we study the  performance of the collective heat engine for moderate finite number of pairs $N$ in the system, and in Sec.\eqref{sec.macroscopic.limit} we explore its properties  in the large $N$ and macroscopic limit  by introducing adequate scalings on the system parameters.
 We present our main conclusions in Sec.\eqref{sec.conclusion}.

\section{Collective Quantum Heat Engine}
\label{sec.models}

\begin{figure}
\includegraphics[width=1\linewidth]{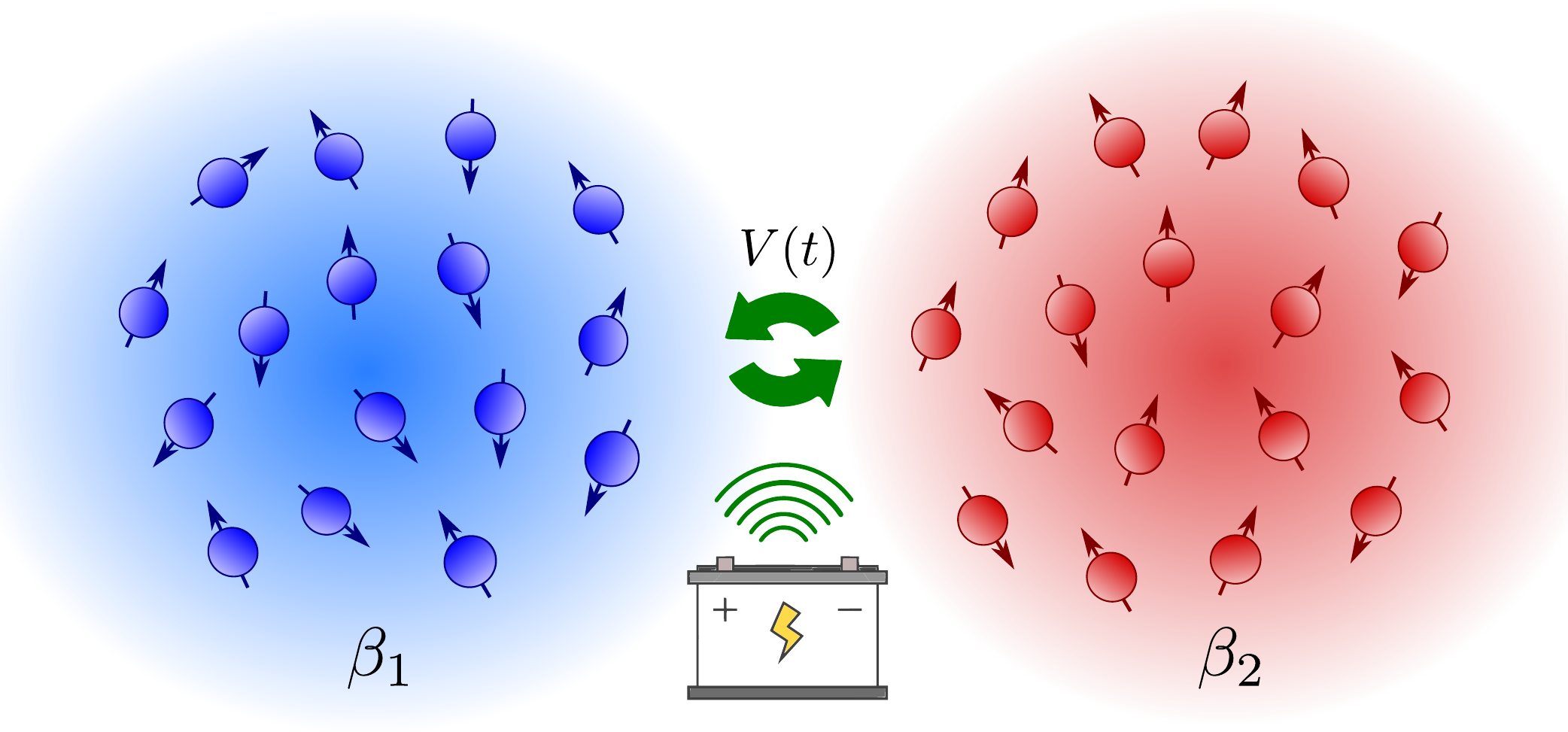}
\caption{Schematic representation of the heat engine composed of two ensembles of
spins (blue and red spins) with different energy spacing, collectively dissipating in respective thermal baths at different temperatures $\beta_1$ and $\beta_2$, and subjected to a collective coherent drive $V(t)$ mediated by a classical field. A heat current from hot to cold baths allows extracting work in the external field (which acts as an ideal battery). Analogously the heat current can be inverted, refrigerating the cold bath, by consuming external work. We explore the 
power, efficiency and constancy of the engine with the number of spins and their enhancements with respect to an equivalent number of two-spin engines working in parallel.}
\label{fig:SchematFigure}
\end{figure}

We consider an engine composed by $N$ pairs of spins, whose inner transitions are collectively driven and coupled to thermal reservoirs, see Fig.~\ref{fig:SchematFigure}. The Hamiltonian of the system $+$ environment is given by,
\begin{equation} \label{eq:total.H}
 \hat H = \hat H_S + \hat H_{SE} + \hat H_E,
\end{equation}
where $\hat H_{S}$ describes the Hamiltonian of the collective, $\hat H_{E}$ is the environment Hamiltonian and $\hat H_{SE}$ the interaction terms between collective system and environment (the details on the environment and interaction Hamiltonians are given in App.~\ref{sec.master}). The Hamiltonian of the system is
\begin{equation}
\hat H_S = \hat H_0 + \hat V(t),
\end{equation}
where $\hat H_0 = \sum_{k=1}^N \hat h_k$ is the sum over the local Hamiltonian of all its $k=1,...,N$ spin-pairs and $\hat V(t)$ is a time-dependent driving term acting collectively over the spins. Each $h_k$ is composed by a pair of $1/2$ spins (or other $2$-level systems) 
Hamiltonian,
\begin{equation}
\hat h_k = E_1 \hat \sigma_{+}^{(1, k)} \hat \sigma_{-}^{(1, k)} + E_2\hat \sigma_{+}^{(2, k)} \hat \sigma_{-}^{(2, k)},
\end{equation}
where $E_{1(2)}$ are the energy splittings (we assume for concreteness $E_2 \geq E_1$), and 
$\hat \sigma_{\pm}^{(i, k)} = (\hat \sigma_{i,k}^{x} \pm i \hat \sigma_{i,k}^{y})/2$ 
 the ladder operator for the $i$'th spin in the $k$'th spin-pair, with $i=1,2$ and $k=1, 2, ..., N$ ($\hat\sigma_{i,k}^{\alpha}$ are the usual Pauli operators).
 
 Collective operators producing simultaneous transitions in the $i$'th spin of all $N$ spin-pairs are operators of the form $\hat S_{i}^{\alpha} = (1/2)\sum_k \hat\sigma_{i,k}^{\alpha}$, for $\alpha = x,y,z$.  The collective raising and lowering spin operators are defined as usual by $\hat{S}_{\pm}^{(i)}=\hat{S}_{i}^{x} \pm i\hat{S}_{i}^{y}$. The collective spin operators inherit the commutation relations from its subsystem components. That is, they satisfy $SU(2)$ commutation relations for the same  $i$'th spin, while commuting otherwise,
$[\hat{S}_{\ell}^{\alpha},\hat{S}_{k}^{\beta}]= i \epsilon_{\alpha\beta\gamma} \hat{S}_{\ell}^{\gamma} \delta_{\ell,k}$,
 with $\epsilon_{\alpha\beta\gamma}$ the Levi-Civita symbol.
 The time-dependent driving term acting collectively in the $N$ pairs reads,
\begin{equation} \label{eq:V}
 \hat V(t) = \frac{\omega_0 }{2}\left(\hat S_{+}^{(1)} \hat S_{-}^{(2)} e^{i \nu t} + \hat S_{-}^{(1)} \hat S_{+}^{(2)} e^{-i \nu t}  \right).
\end{equation}
This term induces a coherent exchange of energy between spins at $E_1$ and $E_2$ and an external classical field at frequency $\nu \equiv (E_2 - E_1)/\hbar$, simultaneously in all the $N$ pairs. In this way $N$ energy quanta $E_2$ in cloud $2$ can be transformed into $N$ quanta $E_1$ in cloud $1$ while augmenting the energy of the classical field by $N \hbar \nu$, and the other way around. Importantly, the collective exchange terms $\hat S_{-}^{(1)} \hat S_{+}^{(2)} e^{-i \nu t} \neq \sum_k \hat\sigma_{-}^{(1, k)} \hat\sigma_{+}^{(2, k)} e^{-i \nu t}$, meaning that they cannot be reduced to 
a simple collection of local transitions between single spin pairs and the drive. 
Since the field is resonant with the energy-gap difference between the two spin clouds, in the interaction picture with respect to $\hat H_0$ the driving term is described by $\hat V_I = (\omega_0/2) \left(\hat S_{-}^{(1)} \hat S_{+}^{(2)} + \hat S_{+}^{(1)} \hat S_{-}^{(2)}\right)$.

 An effective dynamics for the system with $N$ spin-pairs can be derived assuming collective interactions  to the thermal reservoirs.  All spins in cloud $1$ with energy splitting $E_1$ are coupled to a thermal reservoir at inverse temperature $\beta_1$, while all the spin systems in cloud $2$ with splitting $E_2$ are coupled to a second thermal reservoir at inverse temperature $\beta_2 \leq \beta_1$. Therefore, reservoir $1$ will be referred to as 'cold' reservoir, while reservoir $2$ will be the 'hot' reservoir. The inclusion of thermal reservoirs at different temperatures introduces a bias in the direction of the exchanges induced by $V(t)$ in Eq.~\eqref{eq:V}, and depending on the magnitude of the energy splittings $E_1$ and $E_2$, the setup will favor either work extraction in the external field, or refrigeration of the cold reservoir at the expenses of external work consumption, as we will shortly see.
 
 In the interaction picture with respect to $\hat H_0$, one obtains the following GKLS (Lindblad) equation (see App.~\ref{sec.master} for more details) for the dynamics of the system with $N$ spin-pairs (we set $\hbar=1$ through the paper): 
\begin{equation}
\label{eq:masterbtc.pair}
\dot{\hat\rho} = -i[\hat V_I, \hat\rho] + \mathcal{D}_1(\hat\rho) + \mathcal{D}_2(\hat\rho), 
\end{equation}
with $\mathcal{D}_i$ the dissipative part of the Lindbladians generated by the $i$'th thermal reservoir. They read
\begin{eqnarray}
 \mathcal{D}_i(\hat\rho) &=& \gamma_\downarrow^{(i)} \left( \hat S_{-}^{(i)}\hat\rho \hat S_{+}^{(i)} - \frac{1}{2}\{\hat S_{+}^{(i)} \hat S_{-}^{(i)}, \rho \} \right) \nonumber \\
 & & +  \gamma_\uparrow^{(i)} \left(\hat S_{+}^{(i)} \hat\rho \hat S_{-}^{(i)} - \frac{1}{2}\{ \hat S_{-}^{(i)}\hat  S_{+}^{(i)}, \hat \rho \} \right),
\end{eqnarray}
for $i=1,2$.
The rates $\gamma_{\downarrow(\uparrow)}$ of collective emission (absorption) of excitations of cold and hot thermal reservoir are related via the local detailed balance relations
\begin{equation}
 \gamma_\downarrow^{(i)} = \gamma_\uparrow^{(i)} e^{\beta_i E_i}.
\end{equation}
Considering bosonic reservoirs these rates are explicitly written as 
\begin{equation}
\gamma_\downarrow^{(i)}= \Gamma_0 \left(n_\mathrm{th}^{(i)} +1 \right), \quad \gamma_\uparrow^{(i)}=  \Gamma_0  n_\mathrm{th}^{(i)},
\end{equation}
where $\Gamma_0$ is the spontaneous decay rate and $n_\mathrm{th}^{(i)} = 1/\left(e^{\beta_i E_i} - 1 \right)$ is the average number of excitations with energy $E_i$ in the reservoir at inverse temperature $\beta_i$. For fermionic systems we have instead $\gamma_\downarrow^{(i)}= \Gamma_0 \left(1 - f_i \right)$ and  $\gamma_\uparrow^{(i)}= \Gamma_0 f_i$, with $\Gamma_0$ the tunneling rate and $f_i = 1/\left(e^{\beta_i (E_i - \mu_i)} + 1 \right)$ the Fermi distribution. 
We consider throughout this work the case of bosonic reservoirs.

Due to the collective nature of all operators in the model, the system conserves the total spin $S$ for each $i=1,2$. We focus our studies in the case where both collective spins have the same (maximum) total spin $S= N/2$. Physically, the common reservoirs introduce collective excitations of the $N$ spin-$1/2$ particles in their respective ensembles, which are then allowed to be transferred from one ensemble to the other through the driving term $V(t)$ in Eq.~\eqref{eq:V}. The intuition behind this architecture is to profit from superradiance-like effects induced by the common reservoirs to generate an enhanced heat flow between them  that can be used for work extraction.

\section{Steady state operation}
\label{sec.therm.properties}

We are mostly interested in the operation of the heat engine in the long time run, where it reaches a continuous operation mode. We define the steady state of the heat engine ($\hat \rho_{\rm ss}$) from the master equation~\eqref{eq:masterbtc.pair}  through 
\begin{equation}
 -i[\hat{V}_I, \hat{\rho}_\mathrm{ss}] + \mathcal{D}_1(\hat{\rho}_\mathrm{ss}) + \mathcal{D}_2(\hat{\rho}_\mathrm{ss}) = 0.
\end{equation}
In the following we introduce the main concepts characterizing the operation and performance of quantum heat engines working in nonequilibrium steady states, that we will then use in the forthcoming sections to obtain our main results.

In steady state conditions the energy of the system becomes, on average, constant over time. However its nonequilibrium nature allows to establish non-zero heat currents flowing through the system together with a non-zero average power output~\cite{alicki1979,kosloff2014}.
We identify the average power output of the heat engine with the overall change in energy of the machine and thermal reservoirs (introduced by the external classical field). Since the global dynamics of system and reservoirs is closed, this amounts to the evaluation of the time derivative of the total Hamiltonian in Eq.~\eqref{eq:total.H}, whose only time-dependent contribution comes from $\hat V(t)$. In the interaction picture the power output can then be calculated using the operator $\hat P_N(t) = (d\hat{V}(t)/dt)_I$,
whose expectation value for the steady state is given by
\begin{align}\label{eq:power.out}
\mathcal{P}_N &\equiv -\tr[\Big(\frac{d}{d t} \hat{H}(t)\Big)_I ~\hat{\rho}_\mathrm{ss}]= -\tr[ \hat{P}_N(t)6 ~\hat{\rho}_\mathrm{ss}] \\
              &= -i \omega_0~ (E_2 - E_1) \left( \langle \hat S_{+}^{(1)} \hat S_{-}^{(2)} \rangle_\mathrm{ss} - \langle \hat S_{-}^{(1)} \hat S_{+}^{(2)} \rangle_\mathrm{ss}  \right)/2. \nonumber
\end{align}
where the subscript $I$ in the parenthesis is used to denote the interaction picture of the operator $d\hat{H}(t)/dt$ [notice that the derivative of $\hat V(t)$  in the interaction picture is not the derivative of $\hat V_I(t)$] and the minus sign stands from the fact that we are defining the power output exerted by the heat engine (stored in the external field). The second line is obtained explicitly evaluating the derivative of $V(t)$, where expectation values in the steady state are denoted as $\langle \cdot \rangle_\mathrm{ss} \equiv \tr[(\cdot) \rho_\mathrm{ss}]$. The variance of the power output in the steady state can be calculated from~\cite{Molmer2020},
\begin{equation}
\label{eq:power.fluctuation}
	\mathrm{Var}\left(\mathcal{P}_{N}\right) = 
	2 \int_0^\infty dt \left[ \langle \hat{P}_{N}(0) \hat{P}_{N}(t) \rangle_{\rm ss} -  \mathcal{P}_{N}^2  \right],
\end{equation}
 where the two-times correlation function $\langle \hat{P}_{N}(0) \hat{P}_{N}(t) \rangle_{\rm ss} = \langle \hat{P}_{N}(t) \hat{P}_{N}(0) \rangle_{\rm ss}$ can be obtained from the master equation~\eqref{eq:masterbtc.pair} by applying the regression theorem~\cite{breuer2007}. The above expression gives equivalent results to the use of a full-counting statistics approach, as the one used e.g. in Ref.~\cite{kalaee2021violating}.

The average heat currents from the hot and cold heat reservoirs can be calculated from the respective Lindbladians as
\begin{align} \label{eq.heat.flux}
 \dot{Q}_N^{(i)} &\equiv \tr[\hat H_0 \mathcal{D}_i(\rho_\mathrm{ss})] = \langle \mathcal{D}_i^\dagger(H_0) \rangle_\mathrm{ss} \nonumber \\
           &=E_i \Gamma_0 \Big(\langle \hat{S}_{+}^{(i)} \hat{S}_{-}^{(i)} \rangle_\mathrm{ss} + 2n_\mathrm{th}^{(i)} \langle \hat S_{i}^{z} \rangle_\mathrm{ss} \Big),
 \end{align}
where in the first line we used the cyclic property of the trace to obtain the dual for the dissipative Lindbladians $\mathcal{D}_{i}^{\dagger}(\cdot) = \gamma_\downarrow^{(i)}(\hat{S}_{+}^{(i)}(\cdot)\hat{S}_{-}^{(i)} - \{\hat{S}_{+}^{(i)} \hat{S}_{-}^{(i)}, \cdot \}/2) + \gamma_\uparrow^{(i)}(\hat{S}_{-}^{(i)}(\cdot)\hat{S}_{+}^{(i)} - \{\hat{S}_{-}^{(i)} \hat{S}_{+}^{(i)}, \cdot \}/2)$, acting on operators. Notice that in the above definition we didn't include the driving Hamiltonian, $V_I$, which would lead to an extra term in the heat currents of order $\omega_0 \Gamma_0$. Such term needs to be neglected in the weak-driving and weak-coupling regime adopted here in accordance with the approximations taken in the derivation of the master equation~\cite{manzano2018}.  The variance of the heat currents can be also calculated in a similar way to power~\cite{Molmer2020}.

The first law of thermodynamics ensures energy conservation in the steady state. Since the energy of the system does not change on average, we have
\begin{equation} \label{eq:firstlaw}
\mathcal{P}_N = \dot{Q}_N^{(1)} + \dot{Q}_N^{(2)}, 
\end{equation}
 that is, any output power of the engine comes from the energy absorbed from the two reservoirs. On the other hand, the second law of thermodynamics manifests in the non-negativity of the total entropy production in system and reservoirs along generic time evolution:
 \begin{equation}
  \Delta S_\mathrm{tot} = \Delta S - \beta_1 Q_N^{(1)} - \beta_2 Q_N^{(2)} \geq 0,
 \end{equation}
where $\Delta S$ is the change in von Neumann entropy of the engine and $Q_N^{(i)}$ are the integrated heat currents from the reservoirs. Since the evolution is Markovian, and the changes in entropy of the systems vanishes in the steady state, the above equation translates in the non-negativity of the entropy production rate~\cite{spohn:2007}:
  \begin{equation} \label{eq:secondlaw}
  \dot{S}_{\rm{tot}} = - \sum_i \beta_i \dot{Q}_N^{(i)} \geq 0,
 \end{equation}
which limits the regimes actually reachable by the engine, and provides universal bounds on the engine efficiency.

The two main modes of operation that we will explore correspond to a heat engine and a refrigerator. In the first case a positive power output $\mathcal{P}_N \geq 0$ is obtained feeded by a heat current from the hot reservoir $\dot{Q}_N^{(2)} \geq 0$. On the other hand, the chiller regime is characterized by a heat current absorbed from the cold reservoir $\dot{Q}_N^{(1)} \geq 0$, that consumes power from the external driving field $\mathcal{P}_N \leq 0$. The efficiencies of these two modes of operation can be respectively defined by the ratio of the corresponding output useful current to the input source as:
\begin{equation}
 \eta \equiv \frac{\mathcal{P}_N}{\dot{Q}_N^{(2)}} ~~;~~ \epsilon \equiv \frac{\dot{Q}_N^{(1)}}{-\mathcal{P}_N},
\end{equation}
where $\epsilon$ is usually referred to as the coefficient of performance (COP). 

Exploiting the first law in Eq.~\eqref{eq:firstlaw} we can rewrite Eq.~\eqref{eq:secondlaw} in the two following equivalent ways:
 \begin{align}
  \dot{S}_{\rm{tot}} &= - \beta_1 \mathcal{P}_N +  (\beta_1 - \beta_2) \dot{Q}_N^{(2)} \geq 0, \nonumber \\
  \dot{S}_{\rm{tot}} &= - \beta_2 \mathcal{P}_N -  (\beta_1 - \beta_2) \dot{Q}_N^{(1)} \geq 0
 \end{align}
 which lead to Carnot bounds for the efficiency and COP  
 \begin{equation}\label{eq:eff.bounds}
  \eta \leq 1 - \frac{\beta_1}{\beta_2} = \eta_C, ~~;~~
  \epsilon \leq \frac{\beta_2}{\beta_1 - \beta_2} = \epsilon_C,
 \end{equation}
reachable under reversible conditions, when $\dot{S}_\mathrm{tot} \rightarrow 0$. 

Additionally, the TUR imposes extra constraints to the relation between power and efficiency in steady-state heat engines, leading to a trade-off relation which incorporates the fluctuations around the average power~\cite{PhysRevLett.120.190602}:
\begin{equation} \label{eq:TUR}
    \frac{\textrm{Var}\left(\mathcal{P}_{N}\right)}{\mathcal{P}_{N}} \geq \frac{2 \eta T_1}{\eta_C-\eta}.
\end{equation}
This predicts a lower bound on the precision through the power Fano factor $\textrm{Var}\left(\mathcal{P}_{N}\right)/\mathcal{P}_{N}$, which cannot be overcome by any classical Markovian engine working in a non-equilibrium steady state. Based on Eq.~\eqref{eq:TUR}, it is convenient to introduce the normalized constancy of the engine by computing the ratio of the right and left-hand sides~\cite{PhysRevB.98.085425}:
\begin{equation}\label{eq:constancy}
 \mathcal{C}_N \equiv \frac{\mathcal{P}_{N}}{\textrm{Var}\left(\mathcal{P}_{N}\right)} \frac{2\eta  T_1}{\eta_C-\eta},
\end{equation}
which is inversely proportional to the Fano factor of the output power. Since the dynamics of the quantum heat engine we present here is Markovian and has time-independent rates [see Eq.~\eqref{eq:masterbtc.pair}], we have $\mathcal{C}_N \leq 1$ in virtue of Eq.~\eqref{eq:TUR}, whenever it behaves as a classical (stochastic) engine. In this sense, violations of Eq.~\eqref{eq:TUR} in our model, leading to an enhanced constancy $\mathcal{C}_N > 1$, can be interpreted as quantum signatures of the engine.

\section{Enhancements of the performance with $N$}
\label{sec.finite-size}

\begin{figure*}[thb]
 \includegraphics[width=\linewidth]{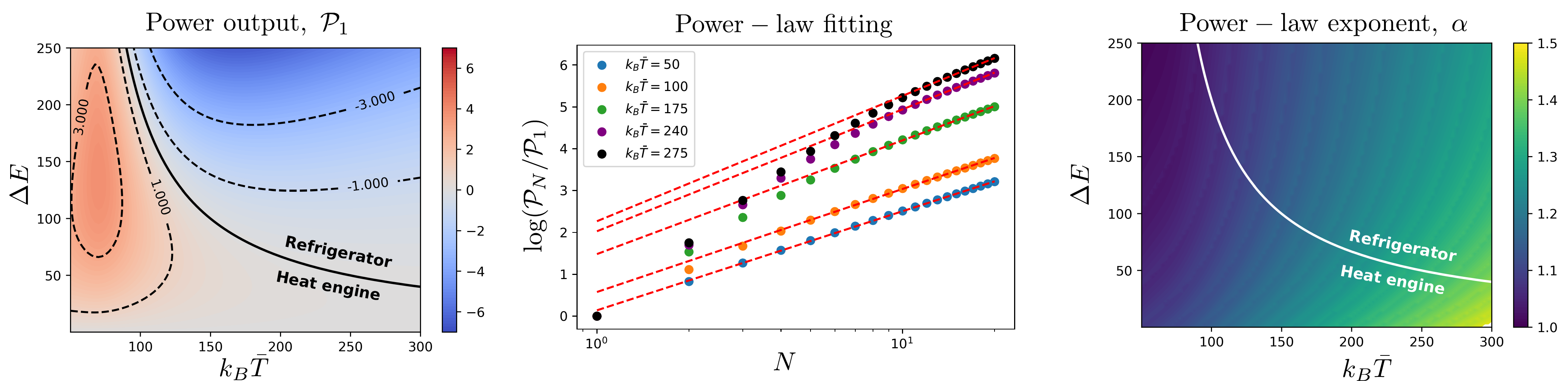}
\caption{{\bf (a)} Power output for a single spin-pair thermal machine, $\mathcal{P}_1$ as a function of energy spacing difference $\Delta E=E_2 - E_1$ and average temperature $\bar{T} \equiv (T_1 + T_2)/2$. The black thick curve, $\Delta E = \Delta E_\ast$, corresponds to zero power where all average currents vanish and highlights the boundary between heat engine (reddish area) and refrigerator (bluish area) modes of operation. {\bf (b)} Example of fittings of the power ratio $\mathcal{P}_N/\mathcal{P}_1 \propto N^{\alpha}$ as a power law with respect to the number $N$ of spin-pairs in the machine, for sufficiently large $N \simeq 20$. Colored circles correspond to simulations for different values of $k_B \bar{T}$ (see legend) and $\Delta E = 125~\omega_0$, while red dashed lines are the result of a linear regression. {\bf (c)} Results for the fitted power-law exponent $\alpha$ as a function of $k_B \bar{T}$ and $\Delta E$. Values of $\alpha > 1$ in extensive regions of the parameter space indicate a super-linear scaling of the collective power $\mathcal{P}_N$ (see color legend on the right). The white thick line corresponds again to $\Delta E= \Delta E_\ast$ in Eq.~\eqref{eq:relation2}. In all plots energetic quantities are given in units of $ \omega_0$. Other parameters are $E1=100~\omega_0, \Gamma_0 = \omega_0, k_B \Delta T = T2-T1= 100~\omega_0$. 
}
\label{fig:regimes}
\end{figure*}

We first explore the performance (power and efficiency) of the collective heat engine introduced above for finite number of pairs $N$. In order to obtain the thermodynamic quantities of interest we numerically solve the steady state of Eq.~\eqref{eq:masterbtc.pair} and evaluate the expressions \eqref{eq:power.out}-\eqref{eq.heat.flux}. The case $N=1$, corresponding to a single pair of spins, can be solved analytically using standard methods~\cite{kosloff2014}. We enforce weak coupling to the reservoirs and weak driving by setting $\omega_0 \sim \sqrt{\Gamma_0} \ll E_1$, while varying $E_2$ and hence $\Delta E = E_2 - E_1$. Fixing the temperature gradient between the thermal baths  $\Delta T \equiv (T_2 - T_1)$, we also explore different regimes by tuning the average temperature $\bar{T} \equiv (T_1 + T_2)/2$.

The average heat currents within the machine are related to the output power through: 
\begin{equation}\label{eq:relation}
\frac{\mathcal{P}_N}{\Delta E} = - \frac{\dot{Q}_N^{\mathrm{(1)}}}{E_1} =  \frac{\dot{Q}_N^{\mathrm{(2)}}}{E_2}.
\end{equation}
This proportionality in the steady state is a consequence of the conservation of the number of excitations between reservoirs, and the fact that every energetic contribution within the setup is associated to a single energy spacing in the thermal machine, that is, $E_1$, $E_2$ or $\Delta E$. This is a characteristic trait of the type of model we are scaling up~\cite{scovil1959,brunner2012}. Relation~\eqref{eq:relation}
 implies in particular that the efficiency and COP coefficient in the current model are given by:
\begin{equation}\label{eq:efficiencies}
 \eta = \frac{\Delta E}{E_2}, ~~~~ \epsilon = \frac{E_1}{\Delta E},
\end{equation}
The above expressions reveal that in both heat engine and refrigerator modes of operation, the efficiency is independent of $N$, and identically equal to the single spin-pair case. The conditions for achieving Carnot efficiency $\eta_C$ (and Carnot COP $\epsilon_C$) are then obtained by combining Eq.~\eqref{eq:eff.bounds} with Eq.~\eqref{eq:efficiencies}, which leads to $E_1/E_2 \equiv T_1 / T_2$. This equilibrium point is achieved when the energy split difference $\Delta E$ or the average temperature $\bar{T}$ equal the following values:
\begin{equation}\label{eq:relation2}
\Delta E_\ast = \frac{E_1 \Delta T}{\bar{T} - \Delta T/2} ~~;~~ \bar{T}_\ast = \frac{\Delta T}{2} \left(\frac{\Delta E - 2 E_1}{\Delta E} \right),  
\end{equation}
for which all average currents simultaneously vanish. The values in Eq.~\eqref{eq:relation2} also determine the boundary between thermodynamic modes of operation of operation in the machine. Similar conclusions were obtained in slightly different models for small heat engines, like the cyclic SWAP model presented in Ref.~\cite{campisi2015}.

The modes of operation of the model are shown in Fig.~\ref{fig:regimes}a, together with the power output $\mathcal{P}_1$ for the $N=1$ case, as a function of the energy spacing difference, $\Delta E$, and the average temperature $\bar{T}$. The machine acts as a heat engine within the reddish area ($\Delta E < \Delta E_\ast$), where positive values of the output power are obtained, accompanied by a heat current from hot to cold reservoirs ($\dot{Q}_1^{(2)} > 0$ and $\dot{Q}_1^{(1)} <0$). By sufficiently increasing the average temperature $\bar{T}$ for fixed gradient $\Delta T$, or by sufficiently increasing the difference in energy spacing ($\Delta E > \Delta E_\ast$) for fixed $E_1$, the engine power reduces and becomes negative (bluish area). In this region the heat currents change sign ($\dot{Q}_1^{(2)} < 0$ and $\dot{Q}_1^{(1)} > 0$) and we obtain a power-driven refrigerator, which consumes input power from the driving to generate a heat flow against the temperature bias.

We are particularly interested in the scaling of the collective output power and their fluctuations (resp. cooling power in the refrigerator regime) with the number of spin-pairs $N$ composing the thermal machine. In Fig.~\ref{fig:regimes}b we show the logarithm of the ratio between the average output power for a $N$-pairs machine, $\mathcal{P}_N$, over the one of a single-pair engine, $\mathcal{P}_1$ as a function of $N$ for different choices of the average temperature $\bar{T}$ and fixed $\Delta E$. The curves for small values of $N$ indicate a highly non-linear behavior of $\mathcal{P}_N$ (note the logarithmic scale), which then gets smoothed to show a linear behavior (see dashed lines) as $N$ increases. This indicates a power law behavior:
\begin{equation}
\mathcal{P}_N /\mathcal{P}_1 \sim N^{\alpha} 
\end{equation} 
for $N$ sufficiently large. We will refer to this regime as the intermediate $N$ regime, which we will compare later on with the macroscopic limit $(N \rightarrow \infty)$ case.

In order to determine the exponent $\alpha$ of the power law, we numerically determine the value of $N$ before the power-law behavior emerge, $N_\mathrm{sat}$, and then perform a linear fit of the numerical curves for using the points $N > N_\mathrm{sat}$. The dashed lines in Fig.~\ref{fig:regimes}b represent examples of the linear fit, the slope of which provide the values of the exponent $\alpha$. The extensive results for the exponent $\alpha$ from our simulation and fitting procedure are shown in Fig.~\ref{fig:regimes}c as a function of $\Delta E$ and $\bar{T}$. They clearly show that the exponent $\alpha$ lies between $1$ and $1.5$ in all the parameter regime studied, implying a super-linear scaling of the collective power $\mathcal{P}_N$  whenever $\alpha > 1$. Similar results are obtained for the collective cooling power $\dot{Q}_N^{(1)}$, which according to Eq.~\eqref{eq:relation} behaves also according to a power law with the same exponent $\alpha$, that is, $\dot{Q}_N^{(1)} = - (\Delta E/E_1) \mathcal{P}_N  \sim N^{\alpha}$.

\begin{figure}[b]
\includegraphics[width = 1 \linewidth]{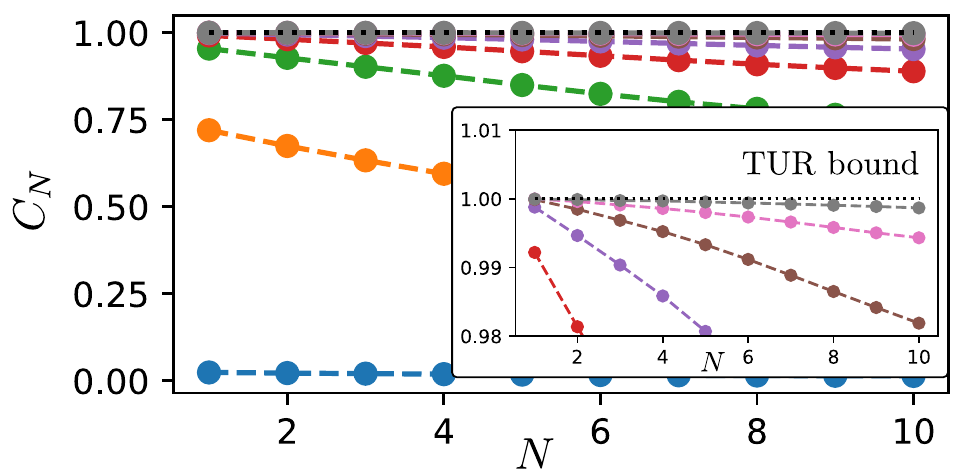}
 \caption{Collective engine constancy $\mathcal{C}_N$ as a function of the number of spin pairs $N$ for different values of the average temperature in the range $k_B \bar{T} = [10 E_1,12 E_1]$ from bottom (blue) to top (grey) curves and TUR bound (black dotted line). Inset: Augmentation of the region close to the TUR bound. Other parameters are $\omega_0 = 0.006 E_1$, $\Gamma_0 = 0.001 E_1$, $\Delta E = 9 E_1$ and $\Delta T = 10 E_1$. The constancy of the engine $\mathcal{C}_N$ decreases with the number of spin pairs $N$ which indicates that the increasing in the power output with the size is accompanied by an even faster growing of the power fluctuations.}
\label{fig:constancy.finitesize}
\end{figure}

 The enhancements in the power output reported above are, however, accompanied by an even faster growing of the power fluctuations as given by its variance. This produces a drop in the normalized constancy of the engine $\mathcal{C}_N$ when we increase $N$, indicating a loose of stability with respect to fluctuations. Even in the regions of parameters where the single spin-pair engine model (slightly) overcomes the classical TUR bound, $\mathcal{C}_N \leq 1$, quantum enhancements of the constancy are quickly lost for larger sizes. This behavior is illustrated in Fig.~\ref{fig:constancy.finitesize}, where we show the normalized constancy, as defined in Eq.~\eqref{eq:constancy}, as a function of $N$ for different values of $k_B \bar{T}$ and fixed $\Delta E = 9 E_1$. For some of the selected set of parameters the single-pair engine achieves values of the constancy around $\mathcal{C}_N \simeq 1.0001$ (grey upper line for $N=1$). However these quantum enhancements of the constancy are lost for larger sizes, since the constancy drops below the classical limit by just considering $N \geq 2$.

Some important remarks concerning the  extrapolation of our results to the macroscopic limit are worth mentioning  at this point.  Even though we have obtained a super-linear scalings for moderate finite system sizes, it does not imply that it shall persists in the macroscopic limit. There might be a crossover size $N_{\rm cr}$ depending on the system parameters, such that a rather different scaling behavior is dominant for sufficiently large system sizes $N \geq N_{\rm cr}$ ---as observed in some recent works in the literature \cite{niedenzu2018,latune2020,watanabe2020} for different models--- where the collective enhancements become suppressed. Moreover, it is important to notice that the absence of a well defined macroscopic limit in the collective Hamiltonian might be the source of spurious superextensive scalings, as it has been recently shown for the charging power in some many-body models of quantum batteries~\cite{PhysRevLett.125.236402,PhysRevResearch.2.023113}. This fact might not only affect the scaling of the engine average power with $N$, but also of their  fluctuations.
 
Furthermore, our analysis for the thermodynamic properties in finite system sizes focused specifically in the steady state of the Lindbladian, \textit{i.e}, the zero eigenvalue solution to Eq.~\eqref{eq:masterbtc.pair}, $\mathcal{L}[\rho_{\rm ss}] = 0$. The spectral properties of the Lindbladian, which contain relevant information concerning the relaxation times 
towards the steady state 
were not taken in consideration so far.  
In order to extrapolate our results to a macroscopic limit the 
properties of the spectral gap of the Lindbladian 
should be studied in connection, due to its physical implications. In particular, for collective models one may have situations in which the macroscopic limit ($\lim_{N \rightarrow \infty})$ and the steady state limit ($\lim_{t \rightarrow \infty}$) do not commute, making even more intricate the analysis of the steady state properties in the macroscopic limit and possible symmetry breaking phases.

Therefore, in order to properly define the macroscopic limit, one may need to scale adequately some of 
the system parameters, like the many-body driving term $V(t)$, or normalize the Lindbladian collective relaxation rates $\gamma_{\uparrow \downarrow}$. 
We deal with this issues
in the next section and introduce a specific set of scaling parameters in a high-temperature regime that guarantee a well behaved macroscopic limit. Notably, this limit can be accessed semi-analytically within a third cumulant approach, and the introduction of the scalings modify the performance and stability of the heat engine with respect to fluctuations.
  
\section{Engine performance in the macroscopic limit}  
\label{sec.macroscopic.limit}

The scaling of the system parameters with the number of spin-pairs $N$ is crucial for a proper analysis of the system in the macroscopic limit, 
as can be seen from its dynamical equations of motion.
 Using the cyclic property of the trace, one can write the time evolution of any observable $\hat O$ in the Heisenberg picture as follows:
\begin{equation}\label{dodt}
\frac{d \langle \hat{O} \rangle  }{dt}=  
\langle i [\hat V_I,\hat O] \rangle  + \langle \mathcal{D}^\dagger (\hat O) \rangle,
\end{equation}
with $\mathcal{D}^\dagger_i$ the dissipative term for the $i$'th thermal reservoir in the Heisenberg picture,
\begin{eqnarray}
 \langle \mathcal{D}^\dagger_i(\hat O) \rangle &=& 
   \frac{\gamma_{\downarrow}^{(i)}}{2} 
 \mathrm{Tr} \left(\left(\left[\hat{S}_{+},\hat{O}\right]\hat{S}_{-}+\hat{S}_{+}\left[\hat{O},\hat{S}_{-}\right]\right)\hat{\rho}\right) \nonumber \\
 & & + \frac{\gamma_{\uparrow}^{(i)}}{2}
 \mathrm{Tr} \left(\left(\left[\hat{S}_{-},\hat{O}\right]\hat{S}_{+}+\hat{S}_{-}\left[\hat{O},\hat{S}_{+}\right]\right)\hat{\rho}\right).\nonumber \\ 
 & &
\end{eqnarray}
It is convenient to define normalized observables $\hat m^\alpha = \hat S^\alpha/S,$ with $\alpha = x,y,z$, in order to study the system in the macroscopic limit. 
Considering $p$-body correlation observables 
$\hat O \equiv \hat m^{\alpha_1} ... m^{\alpha_p}$
with $\alpha_i = x,y,z$, we notice that their corresponding dynamical Heisenberg equations are of order $\mathcal{O}(N)$ (right-hand side 
of Eq.~\eqref{dodt}) or less for a few specific cases, up to their multiplicative factors $\omega_o$ and $\gamma^{(i)}_{\uparrow/\downarrow}$. These dynamical equations thus do not have a well defined macroscopic limit since its various terms do not scale in the same way with the number of spins $N$.
 A possible form to deal with such issue could be a simple renormalization of the multiplicative factors $\omega_o$ and $\gamma^{(i)}_{\uparrow/\downarrow}$ with system size.

Furthermore, looking explicitly for the dissipative   
contribution on the observables, as \textit{e.g.}  its collective magnetization, we obtain
\begin{eqnarray}\label{eq.diss1}
 \langle \mathcal{D}^\dagger( \hat{m}^{\alpha} )  \rangle & = & \frac{N \, \Gamma_0 }{2}\left[ \Re( \langle \hat{m}^{\alpha}\hat{m}^{z}\rangle)  - 
\frac{ (n_\mathrm{th} + \frac{1}{2} )}{S} \langle\hat{m}^{\alpha}\rangle \right], \nonumber \\
 \langle \mathcal{D}^\dagger(\hat{m}^{z}) \rangle & = & - \frac{N \, \Gamma_0 }{2} \left[  1 -\langle(\hat{m}^{z})^2\rangle  +
\frac{ (2n_\mathrm{th} + 1)}{S} \langle\hat{m}^{z}\rangle \right], \label{eq.diss2} \nonumber \\
& & 
\end{eqnarray}
for $\alpha=x,y$, where we used the commutation relations of the collective operators and the fact that the system conserves the total spin $(m^{x})^2 +
(m^{y})^2 + (m^{z})^2 = 1$. We  notice that  the average thermal excitation $n_{\rm th}$ must scale nontrivially with system size, otherwise temperature effects (right terms in the above dynamical equations) are suppressed in the macroscopic limit. In fact, defining the {\textit{high-temperature regime}} as,
 \begin{equation}\label{eq.hight.temp.scaling}
  \omega_0 \rightarrow \omega_0/N,\qquad \Gamma_0 \rightarrow \Gamma_0/N,\qquad \beta \rightarrow  \beta/N,
 \end{equation}
 (where the third term ensure a temperature scaling linearly with the system size) the dissipative dynamical equations reduce to, 
\begin{eqnarray}
 \langle \mathcal{D}^\dagger(\hat{m}^{x(y)}) \rangle & = & \frac{\Gamma_0}{2} \left[ \Re (\langle\hat{m}^{x(y)}\hat{m}^{z}\rangle ) - 
\frac{2 }{\beta E} \langle\hat{m}^{x(y)}\rangle \right],
\label{eq.dyn.eq.diss.mxy}\\
 \langle\mathcal{D}^\dagger(\hat{m}^{z}) \rangle & = & - \frac{\Gamma_0}{2}  \left[ 1 -\langle(\hat{m}^{z})^2 \rangle -
\frac{4}{\beta E} \langle\hat{m}^{z}\rangle \right],
\label{eq.dyn.eq.diss.mz}
\end{eqnarray}
in the macroscopic limit $N\rightarrow\infty$, where we see explicitly its normalization and temperature dependence. It is worth mentioning that one can always estimate the finite $N$ properties in such high temperature regime by a reverse scaling of Eq.~\eqref{eq.hight.temp.scaling}.
From now to the rest of this manuscript we implicitly consider the high-temperature scalings of Eq.~\eqref{eq.hight.temp.scaling} to the Lindbladian parameters unless explicitly stated otherwise.

\textit{Third Cumulant Approach.-} In the next sections we explore the thermodynamic properties of our system in the macroscopic limit within a cumulant approach. Specifically, we approximate the correlations in the system at their (symmetric) third order cumulants,
\begin{eqnarray}
 \langle \hat m^\alpha \hat m^\beta \hat m^\gamma \rangle &=& 
 \langle \hat m^\alpha \hat m^\beta \rangle \langle \hat m^\gamma \rangle + 
 \langle \hat m^\alpha \hat m^\gamma \rangle \langle  \hat m^\beta \rangle \\
  & & + 
 \langle \hat m^\beta  \hat m^\gamma \rangle \langle \hat m^\alpha \rangle - 
 2\langle \hat m^\alpha \rangle \langle \hat m^\beta \rangle \langle \hat m^\gamma \rangle \nonumber 
\end{eqnarray}
(not a simple direct factorization of the expectations values) which generates a closed set of dynamical equations for the macroscopic observables, thus amenable for an analysis. 

\subsection{Purely dissipative case} 
\label{subsec.omega0}

\begin{figure}
\includegraphics[width=0.49\linewidth]{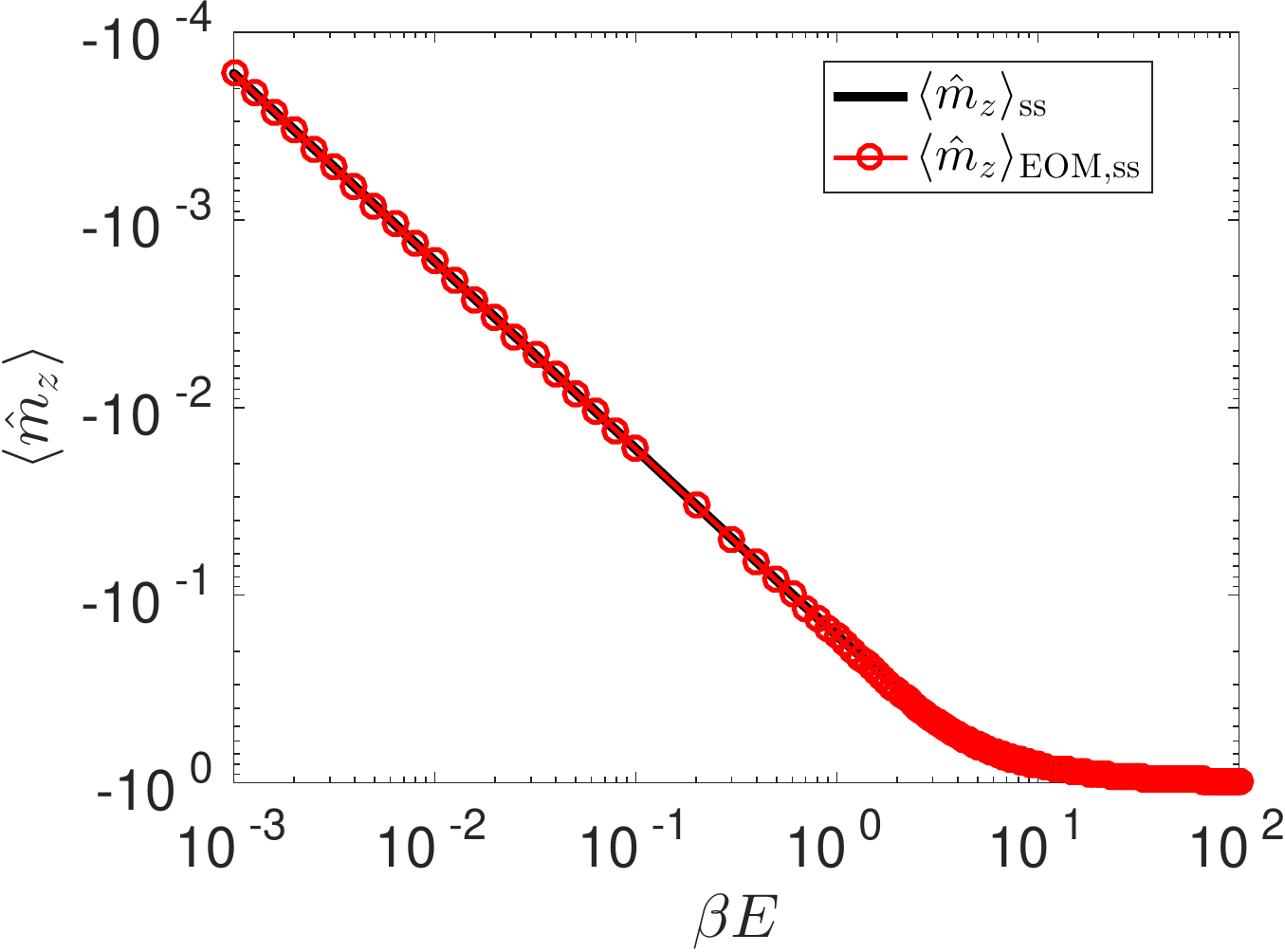}
 \includegraphics[width=0.475\linewidth]{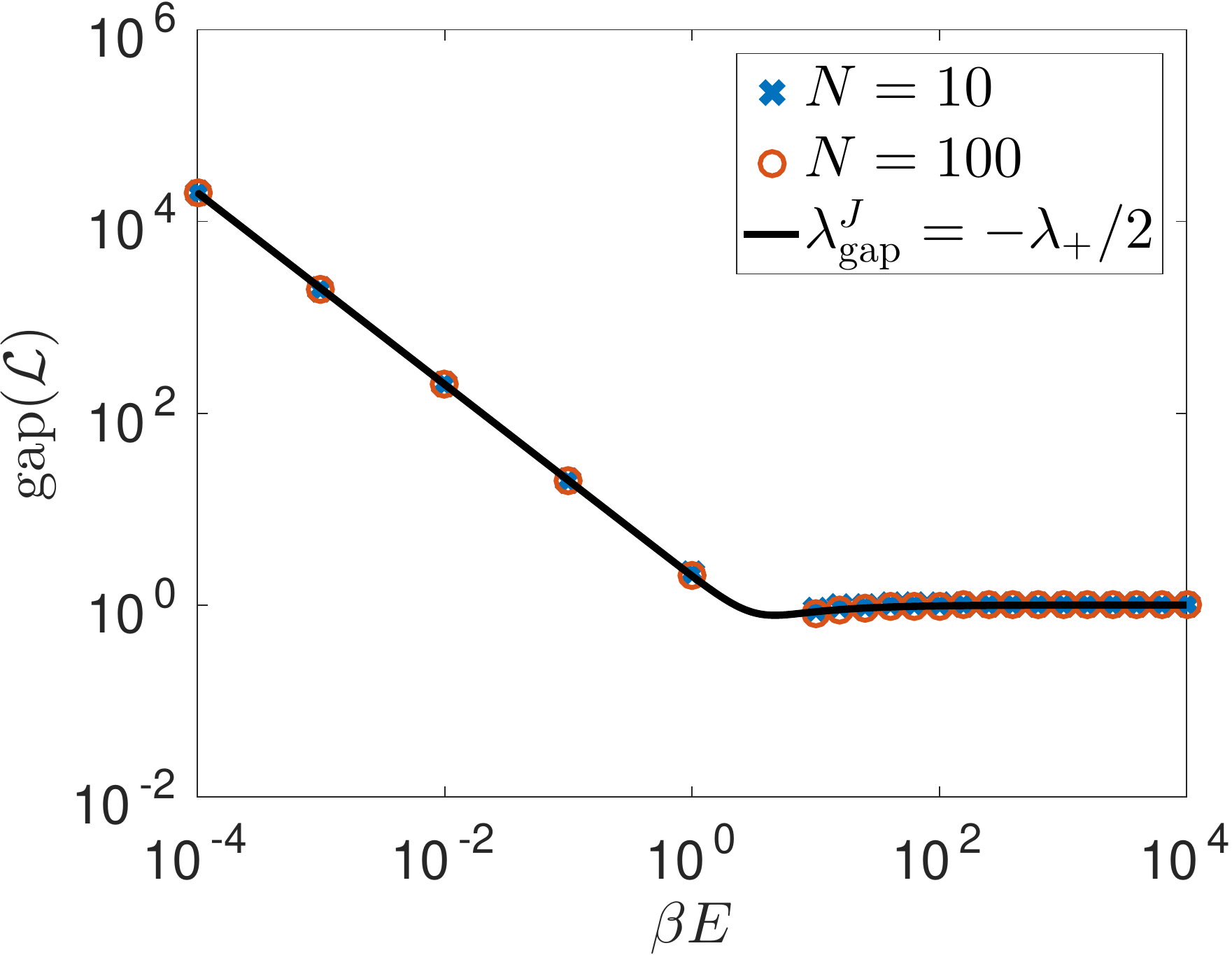}
 \includegraphics[width=0.49\linewidth]{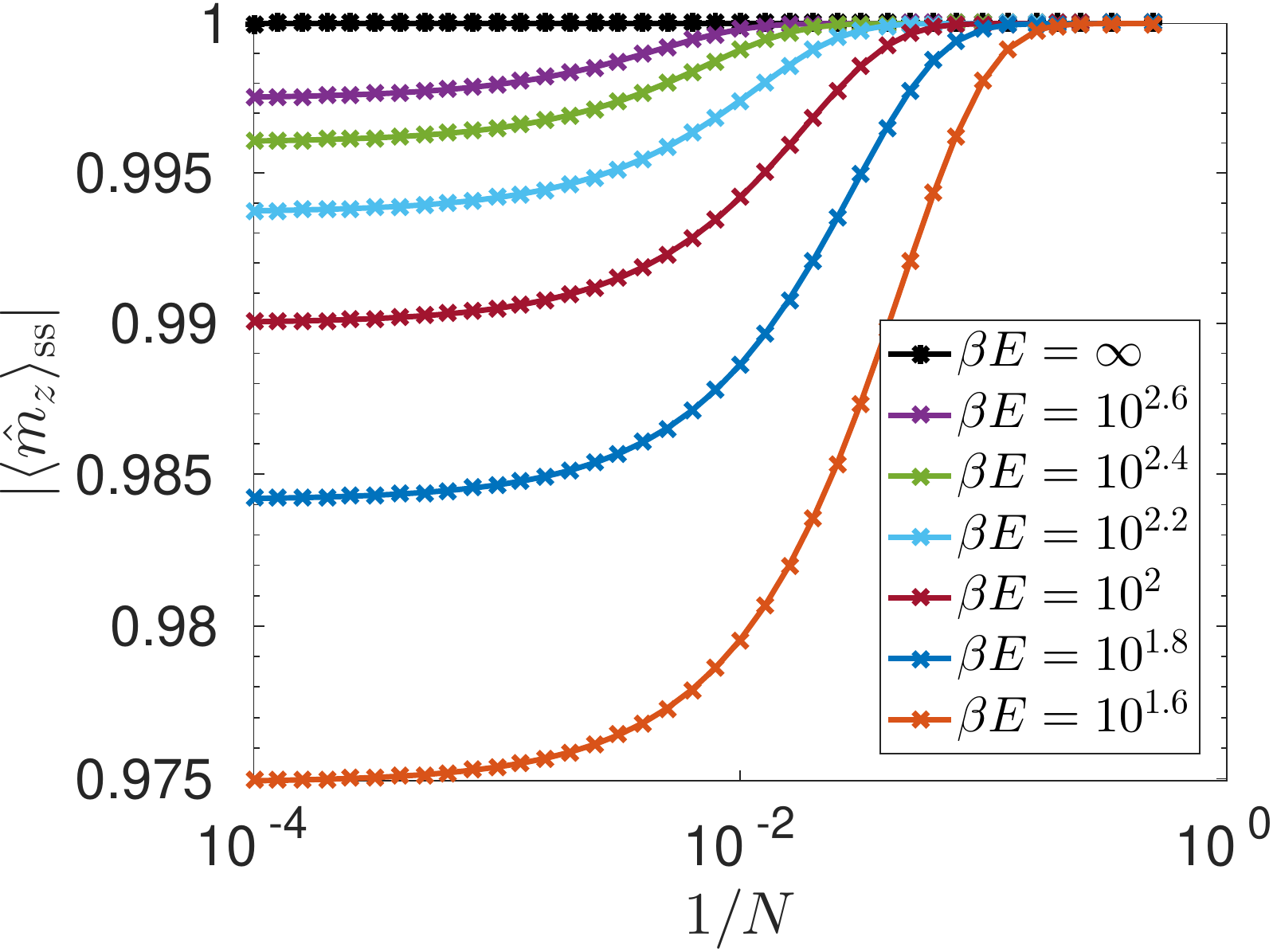}
 \includegraphics[width=0.47\linewidth]{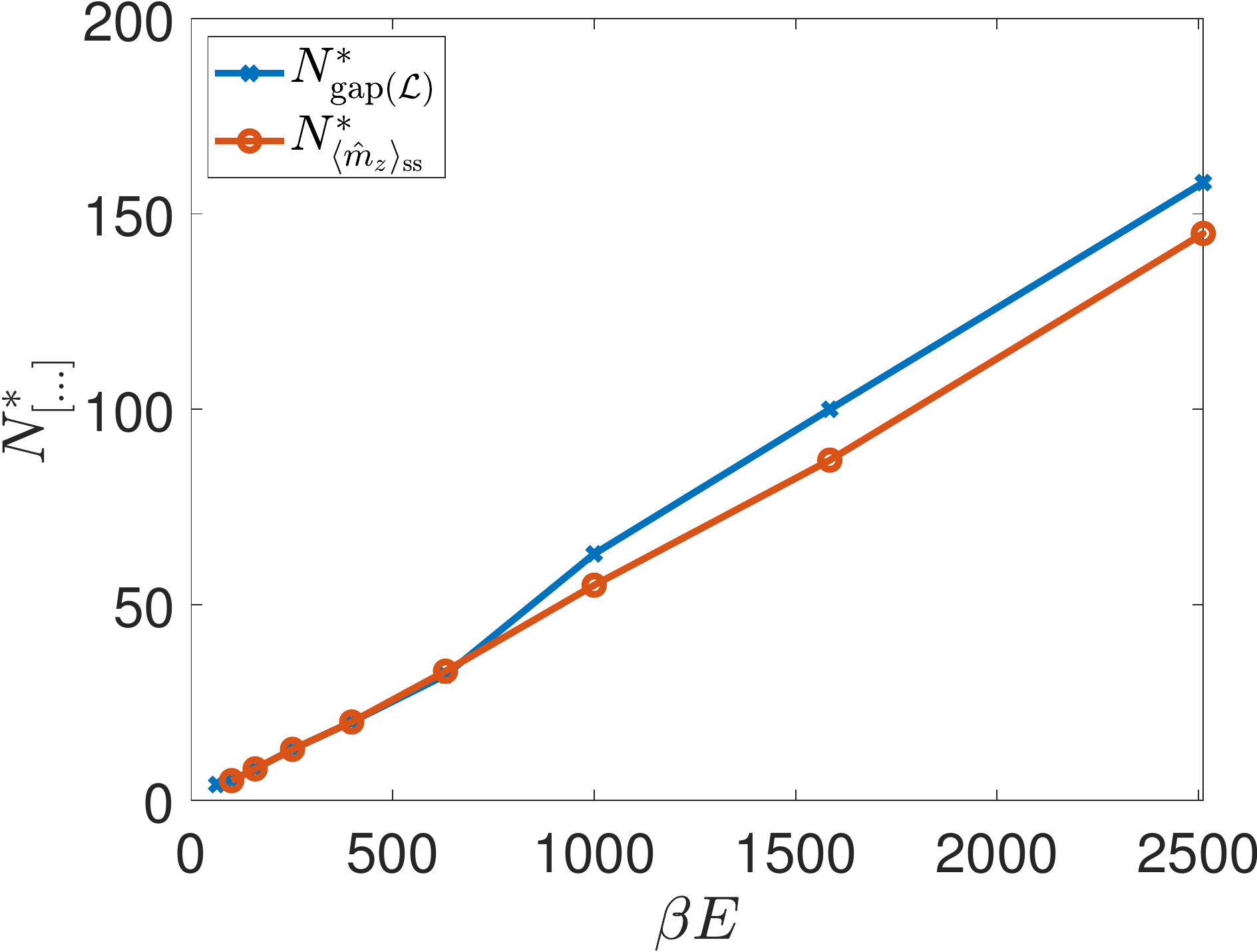}
 \caption{\textit{Purely dissipative case ($\omega_0 = 0$).}  We show in the 
 	top-left panel the steady state magnetization  computed analytically from Eq.~\eqref{eq.ss.magz} and those obtained from the dynamical equations of motion - Eqs.~\eqref{eq:dyn.eq.3rd.dissipation1}-\eqref{eq:dyn.eq.3rd.dissipation2}. Both methods agree with great precision.
 	In the top-right panel we show the Lindbladian
 	gap for finite system sizes, as well the Jacobian gap in the macroscopic limit.
 	In the bottom-left panel we show the steady state magnetization for varying bath's temperatures and system sizes. 
 	In the bottom-right panel we show the transient system size $N^*$ obtained from the Lindbladian gap and from the steady state magnetization, using $\epsilon = 10^{-10}$ for their computation. For sufficient small system size, $N \ll \beta E$, the Lindbladian gap and steady state magnetization are practically indistinguishable from the zero temperature case until reach the transient system size $N^*$.
 }
\label{fig:coll.thermal.bath}
\end{figure}

We first consider the simpler (but far from trivial) case of a purely dissipative Lindbaldian, with $\omega_0 = 0$ in Eq.~\eqref{eq:masterbtc.pair}, in order to highlight the effects arising purely from the collective coupling with the bath, and leave the analysis with the presence of a coherent Hamiltonian and the performance of the quantum heat engine to Subsection~\eqref{sec.pair.2level}. Since in the case of a purely dissipative Lindbladian the two spins in each spin-pair are decoupled, we study the dissipative Lindbladian for a single $i=1$ or $2$, dropping its index from the notations, for convenience.
 We show analytically the steady states of the system, 
 for finite system sizes and in the macroscopic limit, as well as obtain effective dynamical equations which capture with a good accuracy the exact dynamics of the system, thus allowing us to obtain an analytical expression for the Lindbaldian gap.

\textit{Steady states.- } In the purely dissipative case there is no creation of coherence during the dynamics, i.e. there are only decay or excitation jump operators in the Lindbladian. In this way, considering for simplicity an initial state in the diagonal basis of the $\hat m^z$ operator ($\langle \hat m^{x,y}(t=0) \rangle = 0$) it shall remain diagonal during the entire dynamics. The coherences $\langle\hat m^{x(y)}(t) \rangle$ are thus trivial, with
$\mathcal{D}^\dagger ( \hat{m}^{x,y}) = 0$ in Eq.~\eqref{eq.dyn.eq.diss.mxy}. The steady solution of Eq.~\eqref{eq.dyn.eq.diss.mz} is then obtained by the following density matrix,
\begin{equation}
 \hat \rho_{ss} = \mathcal{N} e^{-\frac{\beta E}{2} \hat m^{z} },
 \label{eq.ss.purely.dissipative}
\end{equation}
where $\mathcal{N}=1/\mathrm{Tr}[\mathrm{exp}(-\beta E\hat m^{z} /2)]$ is the normalization constant for the density matrix. The steady state corresponds to a thermal state in the basis of the collective magnetization $\hat m^z$.
For such steady state we can compute its observables analytically (see Fig.~\ref{fig:coll.thermal.bath}-upper left panel):
\begin{eqnarray} \label{eq.ss.magz}
 \langle \hat m^{z} \rangle_{\mathrm{ss}} &=& -\mathrm{coth}\left(\frac{\beta E}{2}\right) +\frac{2}{\beta E}, 
 \label{eq:analytic.steady.state.mag.z} \\
 \langle \hat (m^{z})^2 \rangle_{\mathrm{ss}} &=& 
 1 + 4\frac{\langle \hat m^{z}\rangle_{ss}}{\beta E}, 
 \label{eq:analytic.steady.state.mag.z2} \\
 \langle (\hat{m}^{x})^{2}\rangle_{\mathrm{ss}} & =&\langle (\hat{m}^{y})^{2}\rangle = 2\frac{\mathrm{coth} (\frac{\beta E}{2}) }{\beta E}-\frac{4}{\left(\beta E\right)^2}.
 \label{eq:analytic.steady.state.mag.x2}
\end{eqnarray}

\textit{Time evolution.- } In order to study the dynamics of the system we derive effective dynamical equations closing the expectation values at the $3$rd order cumulant. 
The Heisenberg equation of motion for $\hat S_z^n$ is first derived, obtaining
\begin{eqnarray}
\frac{d (\hat{S}^{z})^n }{dt} &=& 
(\gamma_\uparrow - \gamma_\downarrow) (\hat S^2 - (\hat S^{z})^2) (\hat S^{z})^{n-1} n 
 - n (\hat S^{z})^{n} (\gamma_\uparrow + \gamma_\downarrow) + \nonumber \\ 
 & &
 + (\gamma_\uparrow + \gamma_\downarrow) 
(\hat S^2 - (\hat S^{z})^2) (\hat S^{z})^{n-1} \binom{n}{2}
 +
  \mathcal{O}(S^{<n}),
\end{eqnarray}
which performing the macroscopic limit $S\rightarrow \infty$ and describing in terms of the macroscopic observables $\hat m^\alpha$ reduces to
\begin{eqnarray}
 \frac{2}{\Gamma_0} \frac{d (\hat{m}^{z})^n }{dt} &=& 
- n (\hat m^{z})^{n-1} + n (\hat m^{z})^{n+1} +  \nonumber \\ & &
+ \frac{4}{\beta E} \left\{ \binom{n}{2} (\hat m^{z})^{n-2}
- \left[\binom{n}{2} + n\right] (\hat m^{z})^{n}  \right\}. \nonumber \\ & &
\end{eqnarray}

Closing at the $3$rd cumulant 
$\langle (\hat m^{z})^3 \rangle = 3 \langle (\hat m^{z})^2 \rangle \langle \hat m^{z} \rangle - 2\langle m^{z} \rangle^3$ we obtain the following effective dynamical equations of motion:
\begin{eqnarray}
\frac{2}{\Gamma_0} \frac{d\langle\hat{m}^{z}\rangle}{dt} &=& 
 -(1- \langle (\hat m^{z})^2 \rangle)
 - \frac{4}{\beta E} \langle \hat m^{z} \rangle,
 \label{eq:dyn.eq.3rd.dissipation1} \\
\frac{1}{\Gamma_0} \frac{d\langle(\hat{m}^{z})^2\rangle}{dt}&=& 
\langle \hat m^{z} \rangle (-1 + 3\langle (\hat m^{z})^2 \rangle
- 2\langle \hat m^{z} \rangle^2 ) \nonumber \\
& & + \frac{2}{\beta E}(1-3\langle (\hat m^{z})^2 \rangle).
\label{eq:dyn.eq.3rd.dissipation2}
\end{eqnarray}
We show in Fig.~\ref{fig:coll.thermal.bath}-upper left panel our results for the steady states obtained from the dynamical equations of motion (see Appendix \eqref{sec.appendix.collective.bath} for a more thorough discussion on the effective dynamics). 
We see an accurate agreement compared to the analytical steady state results.

 \textit{Spectral properties of the Lindbladian:} The spectral properties of the Lindbladian can provide further information about the system \cite{iemini2018}. The Lindbladian gap in particular has information on the rate of relaxation towards the steady states of the system. While gapped excitations induce a finite time decay towards the steady states of the system, gapless excitations could support periodic orbits of macroscopic observables in the system which persist indefinitely in time, generating a boundary time crystal phase \cite{iemini2018,Prazeres2021}.
 It is therefore important a proper analysis of the Lindbladian gap in our collective thermal bath and its scaling to the macroscopic limit. The gap is defined as, 
 \begin{equation}
\mathrm{gap}(\mathcal{L}) = - \max_j(\Re(\lambda_j)),
\end{equation}
 where $\lambda_j$ are the eigenvalues of the Lindbladian superoperator. The real part of the eigenvalues, which are always nonpositive, represent the decay times for their corresponding subspaces (generalized eigenstates of the Lindbladian), while the imaginary term describe a coherent evolution on such subspaces. 
 
 We can infer the Lindbladian gap in the macroscopic limit from a stability analysis of their dynamical equations. A simple approach is based on the linearization of these equations around their steady states (fixed points), which is effectively described by the
Jacobian matrix \cite{Strogatz2015}. Specifically, for a given set of dynamical equations $d \langle \hat m^\alpha\rangle /dt = f_{\alpha} (\{ \langle \hat m^\beta \rangle \}_{\beta}) $ with $\beta$ denoting the number of variables and $f_\alpha$ a general nonlinear function over the variables, the Jacobian matrix is defined by the matrix 
$\hat J_{\alpha \beta} = \partial f_{\alpha}/\partial \beta$ representing their (first-order) linear corrections. The Jacobian for our system (Eqs.\eqref{eq:dyn.eq.3rd.dissipation1}-\eqref{eq:dyn.eq.3rd.dissipation2}) is given by,
\begin{equation}
     \hat  J = \begin{pmatrix}
   J_{m^z m^z} &J_{m^z (m^z)^2}\\
   J_{(m^z)^2 m^z} &J_{(m^z)^2 (m^z)^2}
    \end{pmatrix},
    \label{eq.jacobian.general}
\end{equation}
with,
\begin{align}
J_{m^z m^z}&= -2\Gamma_{0}/\beta E,\quad
J_{m^z (m^z)^2} = \Gamma_{0}/2,
\nonumber \\
J_{(m^z)^2 m^z} &=  \Gamma_{0}(-1+3 \langle (\hat{m}^{z})^{2}\rangle_{\mathrm{ss}} -6 \langle \hat{m}^{z}\rangle_{\mathrm{ss}}^2),
\nonumber \\
J_{(m^z)^2 (m^z)^2} &=  3\Gamma_0( \langle \hat{m}^{z}\rangle_{\mathrm{ss}}-2/\beta E).
\end{align}
Its eigenvalues computed at the steady states of the system provides information on the stability of such fixed points. An eigenvalue with negative (positive) real part is related to an attractive (repulsive) fixed point, i.e., a stable (unstable) steady state, with the real part of the eigenvalue describing the rate of decay. We find that the steady states of our system are described by stable fixed points. We thus define the Jacobian gap as,
\begin{equation}\label{eq.jacobian.gap}
 \lambda_{\rm gap}^J = -\max_{\{ \lambda_i\}} \Re(\lambda_i),
\end{equation}
with $\lambda_i$ the eigenvalues of $\hat J$ computed at the stable fixed point. Both the Lindbladian as Jacobian gaps thus represent the slowest decaying modes in their corresponding dynamical equations: while the Lindbladian capture the decay of the full density matrix properties, the Jacobian gap focus to a particular set of observables, the macroscopic magnetizations. Nevertheless, the decay rate at the full density matrix level can be similar to its local observables (in general this is indeed the case, apart from specific kinetically constrained systems, as \textit{e.g.} Ref.\cite{Raul2020} where local particle densities display faster dynamics compared to its (global) density matrix purity, due to existence of boundary wall excitations in the system). In our system we find that these two gaps are indeed proportional to each other.

Using the analytical steady state magnetizations of Eqs.\eqref{eq:analytic.steady.state.mag.z}-\eqref{eq:analytic.steady.state.mag.z2}, we obtain that the Jacobian eigenvalues for the collective thermal bath are given by,
  \begin{equation}
  \lambda_{\pm} = \frac{\Gamma_0}{2}\left(-\frac{2}{\beta E}-3 \mathrm{coth}\left(\frac{\beta E}{2}\right)\pm\sqrt{A}\right),
 \end{equation} 
where,
 \begin{equation}
A = 4+\frac{4}{(\beta E)^2}+12\frac{ \mathrm{coth}\left(\frac{\beta E}{2}\right)}{\beta E}-3\mathrm{coth}\left(\frac{\beta E}{2}\right)^2.
\end{equation} 

We show in Fig.~\ref{fig:coll.thermal.bath}-(top right panel) our results for the Lindbladian gap for finite system sizes as well as the Jacobian eigenvalue in the macroscopic limit.  We see that they are related to each other apart from a proportionality constant, and the Lindbladian is always gapped both for finite system sizes as well as in the macroscopic limit. The gap shows two different regimes with the temperature of the bath. While for small temperatures the gap is approximately constant, for larger ones the gap scales linearly with the temperature. Specifically,
\begin{eqnarray}
 \rm{gap}(\mathcal{L}) &\sim& 1, \qquad \qquad \,\,\,\, \text{for } \beta E  \gtrsim 1.  \nonumber \\
 \rm{gap}(\mathcal{L}) &\sim & (\beta E)^{-1}, \qquad \text{for } \beta E  \lesssim 1.  
\end{eqnarray}

\textit{Transient system size:} Due to the temperature scaling (Eq.~\eqref{eq.hight.temp.scaling}) in the high-temperature regime, the average thermal excitations $n_{\rm th}$ scales with the number of spins in the system. In this way for sufficiently small system sizes $N \ll \beta E$ the average thermal excitations could be approximated, by a first order expansion in terms of $(\beta E/N)$, in the form of $n_{\rm th} \approx (\beta E/N)^{-1} \sim 0$. Thus sufficiently small system sizes resemble the case of a zero temperature bath. 
 We compute the Lindbladian gap and steady state magnetization for increasing temperatures and system sizes. We obtain that these two quantities are roughly indistinguishable from the zero temperature case up to a transient system size $N^*$, corroborating our previous arguments (see Fig.~\ref{fig:coll.thermal.bath}-(bottom left panel) for the magnetization results).  Formally defining $N^*$ as the minimum system size for which 
 $|\mathrm{gap}(\mathcal{L})_{\beta E } - \mathrm{gap}(\mathcal{L})_{\beta E \rightarrow \infty}| > \epsilon$, (analogously for the magnetization), with $\epsilon = 10^{-10}$, we obtain that 
 $N^* \sim 0.05\, \beta E$ for small temperatures ($\beta E \gtrsim 1$) while it is negligible for larger ones - see Fig.~\ref{fig:coll.thermal.bath}-(bottom right panel).

\subsection{Full dissipative-driven heat engine} 
\label{sec.pair.2level}


\begin{figure}
 \includegraphics[width=0.9\linewidth]{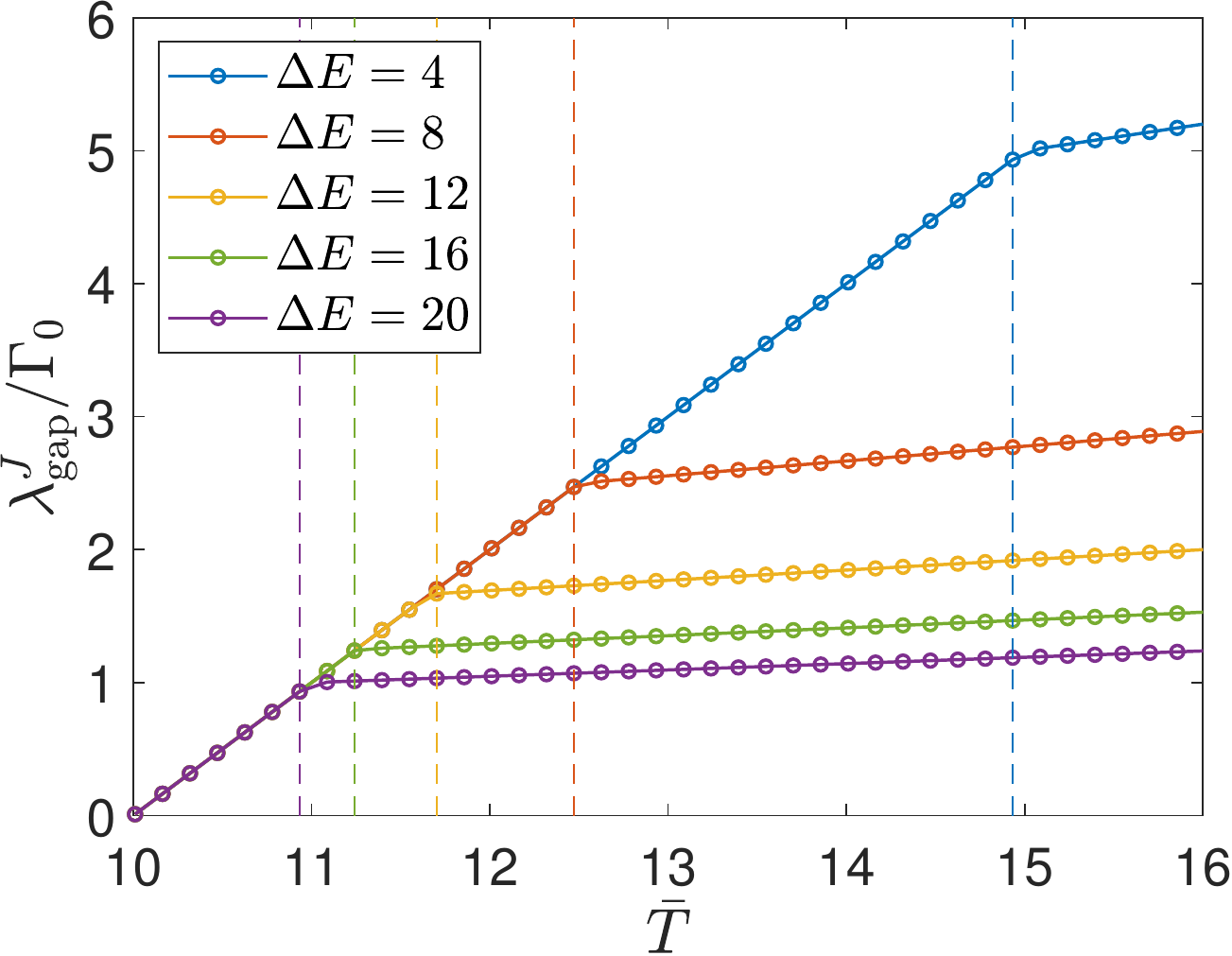}
\caption{Jacobian gap $\lambda_{\rm gap}^J$ for different energy splittings $\Delta E$ and varying average temperatures $\bar{T}$, for a system with coupling parameters $\omega_0 = 0.006, \Gamma_0 = 0.001,\, E_1 = 1,\Delta T = T_2 - T_1 = 20$. The dashed vertical lines highlight the boundary between heat engine (left part) and refrigerator (right part) modes of operation in the system, for the different fixed energy splittings difference. The relaxation to the steady state is faster for higher average temperatures seeing that the Jacobian gap increases linearly with the average temperature, $\Re(\lambda_1^J) \sim c \bar{T}$. In the heat engine mode of operation the slope of the growth is independent of the energy spacing difference which is not true for the refrigerator mode of operation.
}
\label{fig.jac.gap.heatengine}
\end{figure}



\begin{figure*}
 \includegraphics[width=0.32 \linewidth]{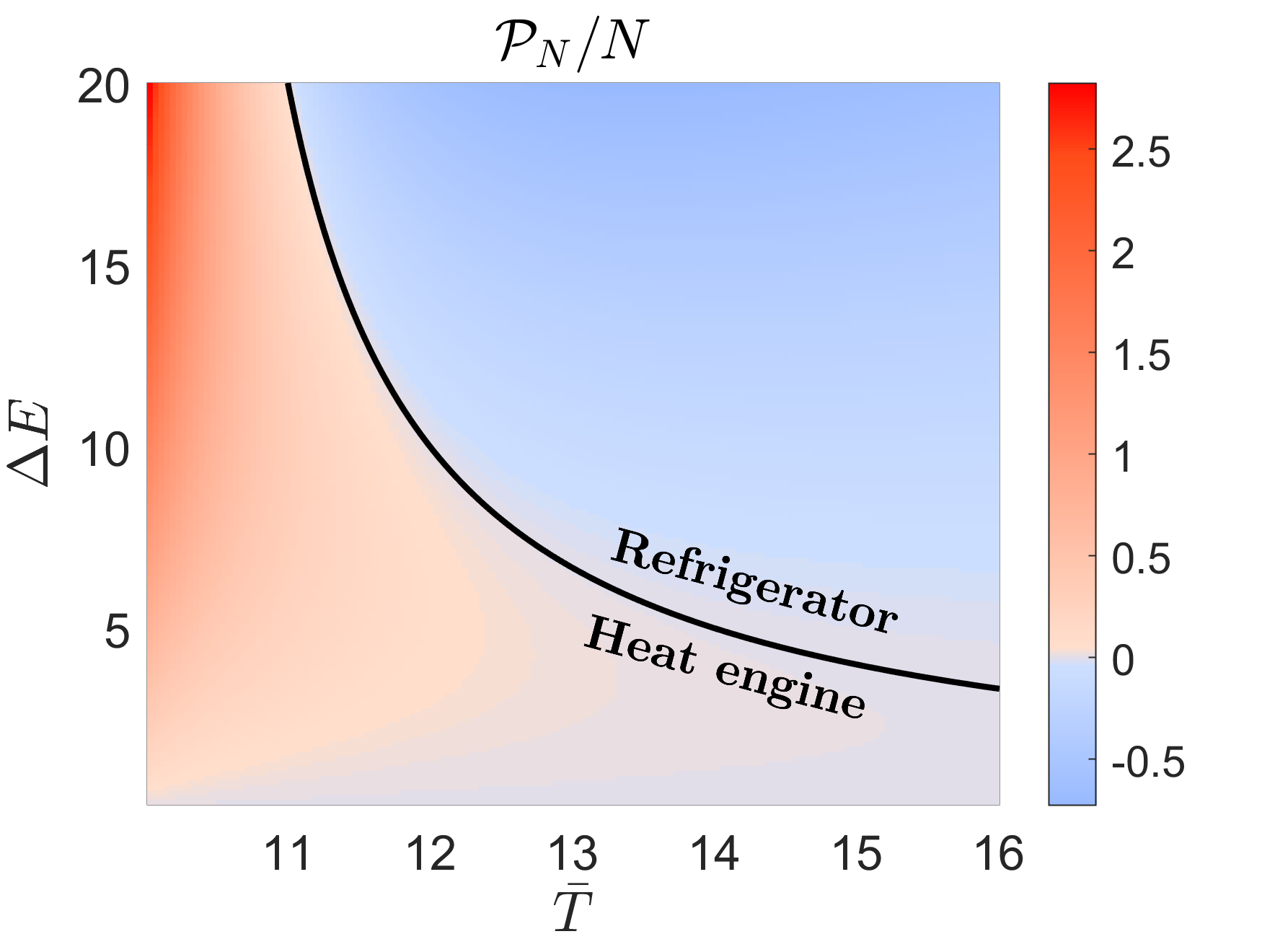}
 \includegraphics[width=0.32 \linewidth]{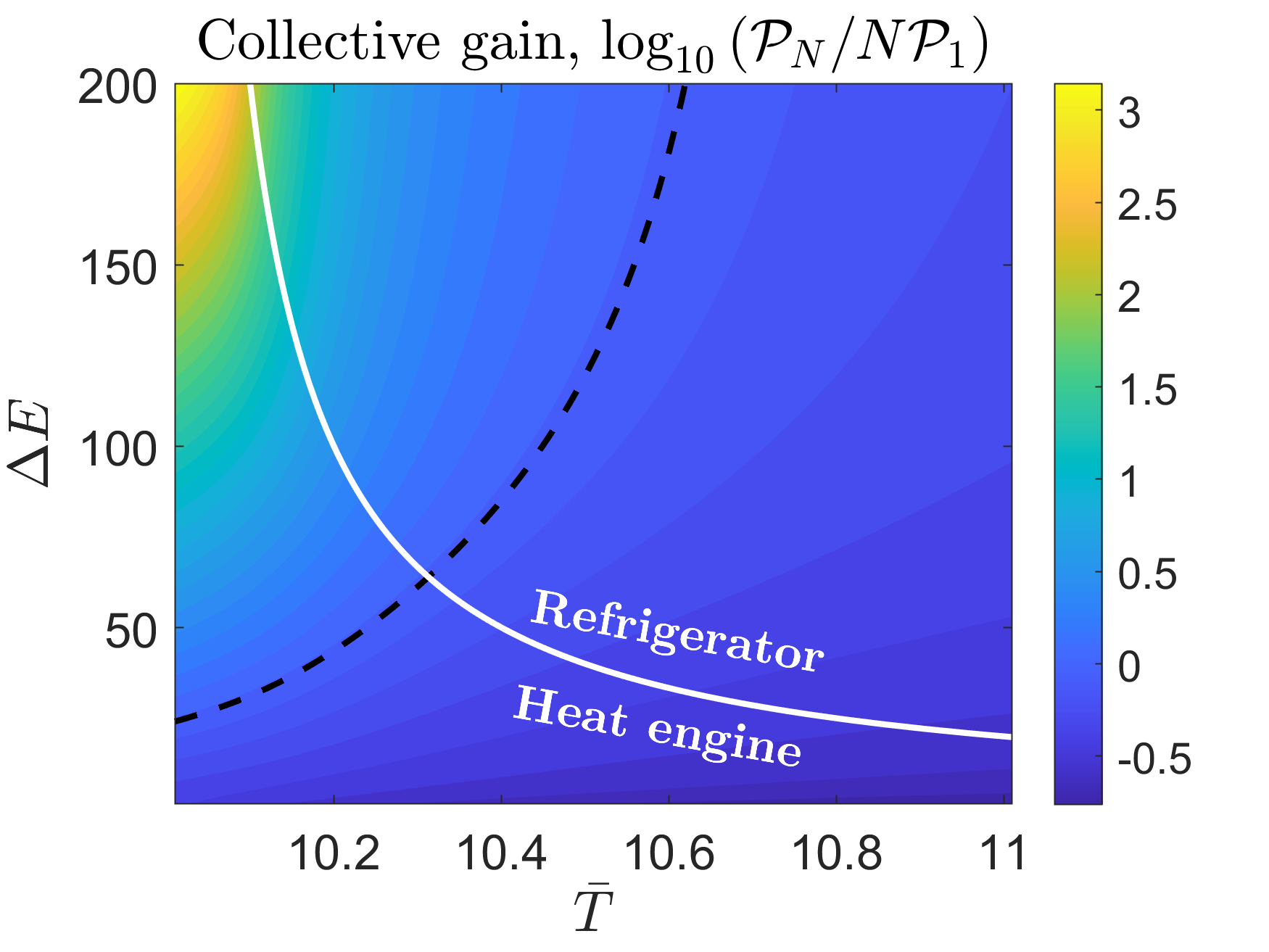}
\includegraphics[width=0.32 \linewidth]{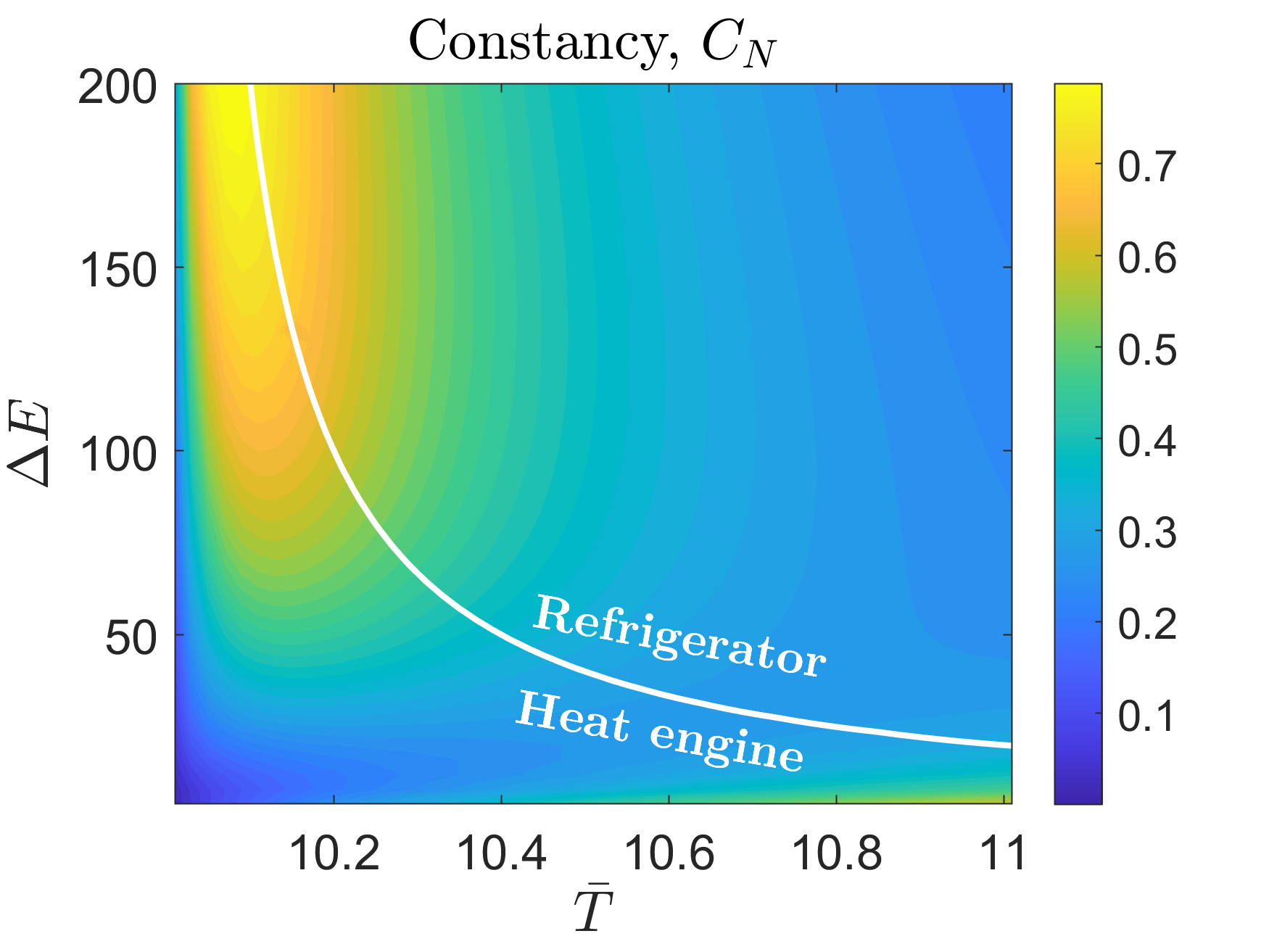}
\caption{ Power, collective gain 
and constancy of the collective heat engine in the macroscopic limit, with fixed parameters 
$\omega_0 = 0.006, \Gamma_0 = 0.001,\, E_1 = 1,\Delta T = T_2 - T_1 = 20$, as a function of the 
energy splitting difference $\Delta E$ and average temperature $\overline{T} = (T_1+T_2)/2$.
{\bf (a)} Output power in units of $\Gamma_0$ for the collective engine, $\mathrm{lim}_{N \rightarrow \infty}(\mathcal{P}_N/N\Gamma_0)$ in the  steady state. The black curve highlights the boundary between heat engine (reddish area) and refrigerator (bluish area) modes of operation, similarly to the single spin-pair case. {\bf (b)} Log-ratio of the collective power output and $N$ independent single-pair engines, $\mathrm{lim}_{N \rightarrow \infty} \log_{10}(\mathcal{P}_N/N \mathcal{P}_1)$. The black dashed curve separates the region with collective gain ($\mathcal{P}_N/N \mathcal{P}_1 > 1$) to the one with collective loss ($\mathcal{P}_N/N \mathcal{P}_1 < 1$).
{\bf (c)} Constancy $\mathcal{C}_N$ of the collective engine working in the steady state, along the same set of system parameters. While the constancy lies below the classical TUR limit, $\mathcal{C}_N \leq 1$, we observe a high stability region ($\mathcal{C}_N \simeq 0.7$) overlaping with high gains in the collective power, $\mathrm{lim}_{N \rightarrow \infty}(\mathcal{P}_N/N \mathcal{P}_1) \simeq 10^3$. 
}
\label{fig.PowerOutput.3rdCum}
\end{figure*}


In this section we include back the coherent driving Hamiltonian in the dynamics ($\omega_0 \neq 0$) and study the performance of the heat engine in the macroscopic limit and high- temperature regime [Eq.~\eqref{eq.hight.temp.scaling}]. Our approach in order to obtain the steady states of the system follows similarly to the previous subsection. The equations of motion are obtained within a third order cumulant approach, from which one can obtain the corresponding steady states through a numerical integration of the dynamics. The dynamical equations for one-body macroscopic observables are given by

\begin{eqnarray}\label{sc.d1x}
\frac{d\langle\hat{m}_{\ell}^{x}\rangle}{dt}&=&\frac{1}{2}\omega_0\langle \hat{m}_{\ell}^{z}\hat{m}_{\bar{\ell}}^{y}\rangle+\frac{\Gamma_0}{2}\Re  \langle \hat{m}_{\ell}^{x}\hat{m}_{\ell}^{z}\rangle  -\frac{\Gamma_{0}}{\beta_\ell E_\ell}\langle\hat{m}_{\ell}^{x}\rangle, \nonumber
\\
\label{sc.d1y}
\frac{d\langle\hat{m}_{\ell}^{y}\rangle}{dt}&=&-\frac{1}{2}\omega_0\langle \hat{m}_{\ell}^{z}\hat{m}_{\bar{\ell}}^{x}\rangle+\frac{\Gamma_0}{2}\Re \langle \hat{m}_{\ell}^{y}\hat{m}_{\ell}^{z}\rangle
-\frac{\Gamma_{0}}{\beta_\ell E_\ell}\langle\hat{m}_{\ell}^{y}\rangle, \nonumber 
\\
\label{sc.d1z}
\frac{d\langle\hat{m}_{\ell}^{z}\rangle}{dt}&=&\frac{1}{2}\omega_0\left(\langle \hat{m}_{\ell}^{y}\hat{m}_{\bar{\ell}}^{x}\rangle-\langle \hat{m}_{\ell}^{x}\hat{m}_{\bar{\ell}}^{y}\rangle\right) + 
\nonumber \\ 
& &\, -\frac{\Gamma_0}{2}\left(\langle \hat{m}_{\ell}^{x}\hat{m}_{\ell}^{x}\rangle+\langle \hat{m}_{\ell}^{y}\hat{m}_{\ell}^{y}\rangle\right)
-\frac{2\Gamma_{0}}{\beta_\ell E_\ell}\langle\hat{m}_{\ell}^{z}\rangle, 
\end{eqnarray}
with $\ell,\bar{\ell} \in \{1,2\}$ and $\ell \neq \bar{\ell}$. The dynamical equations for the two-body observables have a much more complex structure, and we describe them in appendix~\eqref{Appendix.Third.Cumulant.pair}. 

We first study the Jacobian gap $\lambda_{\rm gap}^J$, see Fig.~\ref{fig.jac.gap.heatengine}. As expected from the previous discussion on the purely dissipative case, for increasing average temperatures we have a larger gap, the dynamics towards the steady state thus becomes faster for larger temperatures. We further see an interesting behavior. The gap increases linearly with the average temperature $\Re(\lambda_1^J) \sim c \bar{T}$, where  $c$ is the slope of the growth, and shows two different regimes. For low average temperatures, $c$ is independent on the energy splitting difference $\Delta E$, while for larger average temperatures we have a different growth slope which depends on the energy splittings.  The transition between these two regimes occurs exactly at the transition between the two thermodynamic modes of operation in the machine, from a heat engine to a refrigerator, i.e. at the Carnot point. 
Since our major concern here is the steady-state regime itself, we leave a deeper analysis of this point as an interesting perspective.

 We then focus on the performance of the collective engine in the macroscopic limit. As in the case of finite sizes $N$, the heat currents within the machine and the power output follow the proportionality relation of Eq.~\eqref{eq:relation}, which we checked numerically, and hence the efficiency and COP coefficient of Eq.~\eqref{eq:efficiencies} are still valid. The power output can thus be eventually enhanced in the macroscopic limit without affecting its efficiency. In Fig.~\ref{fig.PowerOutput.3rdCum}a we show the collective power output divided by the number of spin-pairs $N$, in the macroscopic limit and in units of $\Gamma_0$. As can be observed, the boundary between thermodynamic modes of operation (solid black curve) in the machine is also given by Eq.~\eqref{eq:relation2}.

  We notice that with the scalings introduced above [Eq.~\eqref{eq.hight.temp.scaling}] the power output of the quantum heat engine is now bounded by $\mathcal{P}_N \leq \mathcal{O}(N)$ and no super-linear enhancements can be reached within this regime. In any case, we find that one can still have a collective gain compared to the case of independent spins. This is illustrated in Fig.~\ref{fig.PowerOutput.3rdCum}b where we show our results for the total power output of the collective quantum heat engine with $N$ spin-pairs, as compared to the power output of $N$ single-pair engines working in parallel. We observe two regions corresponding to collective gain $\mathcal{P}_N/(N\mathcal{P}_1) > 1$ (upper left) and loss of power $\mathcal{P}_N/(N\mathcal{P}_1) < 1$ (bottom right) as separated by the dashed line. We notice that both regions comprise either the heat engine and refrigerator regimes. For the parameters studied in the figure, larger energy splittings $\Delta E$ and smaller average temperatures $\overline{T}$ (with fixed bias $\Delta T$) lead to higher improvements in the collective power. The critical value for the energy splitting defining the boundary of the collective gain region (dashed line) seems to depend on the average temperature (roughly) algebraically, $(\Delta E)_c \sim (\overline{T})_1^{\mathrm{cte}}$.
 We also obtain that the collective gain has an exponential dependence with the energy splittings, $\mathcal{P}_N / (N \mathcal{P}_1) \propto e^{\Delta E}$ (see appendix~\ref{appendix.power.enhancement.mac}). 
We recall that the regions in the system parameters leading to a collective gain in the power output do not necessarily coincide with the ones where the power output is itself higher, since the gain is a ratio of two power outputs (the collective and the one for $N$ independent pairs of spins). In fact, by comparing Fig.~\ref{fig:regimes}a and Fig.~\ref{fig.PowerOutput.3rdCum}a we observe that, while both collective and individual power outputs have qualitatively the same behavior, these are different from the collective gain represented in Fig.~\ref{fig.PowerOutput.3rdCum}b.

 We also study the constancy $\mathcal{C}_N$ for the system in the macroscopic limit. In order to compute it we need the expectation values up to four-body correlations contained in the power fluctuations, Eq.~\eqref{eq:power.fluctuation}. We compute these correlations within our $3$rd cumulant approach, noticing that once a cumulant order is closed, all of its higher orders are null as well. Therefore, we use the $4$th cumulant closure expression to approximate those four-body correlators using lower orders. The time correlations can then be computed with quantum regression theorem, leading to a simple set of linear dynamical equations which can be solved with standard numerical approaches (see appendix~ \ref{Appendix.power.fluctuations} for more details).

The constancy $\mathcal{C}_N$ in the macroscopic limit is shown in Fig.~\ref{fig.PowerOutput.3rdCum}c. Remarkably, it shows a stronger stability (higher values of $\mathcal{C}_N$) for the engine around the region where we see power enhancements, i.e, small temperatures and large energy splittings. However, we could not see a precise relation between these two engine properties (performance and stability) in general. 
Either way, we notice that the constancy of the system for a macroscopic number of spins lies always below its classical TUR bound, $\mathcal{C}_N \leq 1$. This behavior is expected due to the collective nature of the model: the local properties of the system in the macroscopic limit (such as finite-body correlations) can be described by classical correlations according to the quantum de Finetti theorem~\cite{Watrous2011}.

\begin{figure}
    \includegraphics[width= 1 \linewidth]{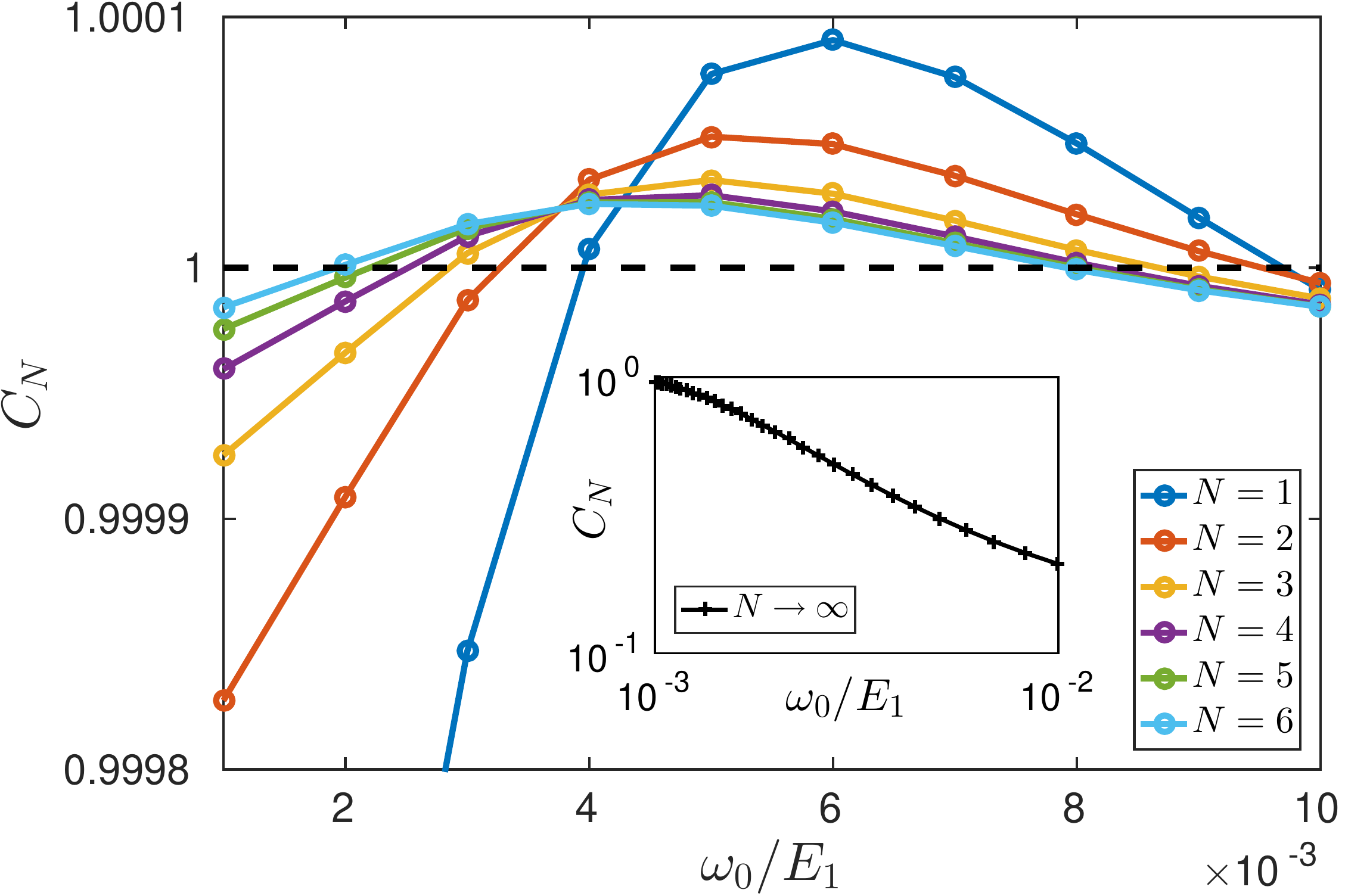}
	\caption{Constancy $\mathcal{C}_N$ for increasing system sizes and in the macroscopic limit (inset panel). We set system parameters $E_1=1.0, E_2=10.0, \Gamma_0=0.001, T_1=2$ and $T_2=25$, for varying $\omega_0$ and considering the system in the high temperature regime. There are regions of parameters with quantum enhancements of the constancy, $\mathcal{C}_N \geq 1$, for large, but finite sizes. These enhancements are lost in the macroscopic limit, where the constancy is always below the classical limit. 
	}
	\label{fig.var.tur}
\end{figure}

Nevertheless, it is interesting to analyse the ``quantum-classical'' crossover for the constancy when increasing the number of spins in system. In Fig.~\ref{fig.var.tur} we show the constancy $\mathcal{C}_N$ as a function of $N$ in the high-temperature regime. We focus specifically in a region of parameters for which $\mathcal{C}_1$ (i.e. the constancy for the case of a single pair of spins) shows violations of the classical TUR bound. We observe that there are regions with enhanced constancy (and an enhanced TUR violation) for larger, but finite sizes (e.g. $\omega/E_1 \sim 0.003$). These enhancements are lost in the macroscopic limit (inset panel), where the constancy is always below one, as discussed above.

It is also worth recalling that the scaling performed in our high-temperature-regime analysis [using Eq.~\eqref{eq.hight.temp.scaling}] should be seen as a tool in order to obtain a well defined macroscopic limit $N\rightarrow \infty$. In the large $N$ limit, however, the analysis of the system without scalings could then be derived by simply reversing the scaling procedure.

\subsection{Mutual Information}

Finally, it is interesting to seek for the roots of the power output in the system in terms of the correlations between its microscopic constituents. We thus study the correlations between two particular spins in the system, each one belonging to a different collective pair.
In order to compute their correlations, we first compute the reduced density matrix for the two spins $\hat \rho_{1,2}$ from a \textit{tomography} procedure: we reconstruct the reduced density matrix using all two-body expectation values $\langle \hat m_1^\alpha \hat m_2^{\alpha'} \rangle$ obtained from the third-cumulant approach developed above. We remark that, within a semiclassical approach (second order cumulant closure), there would not be any correlations between the spins at all. 

\begin{figure}
\includegraphics[width=1 \linewidth]{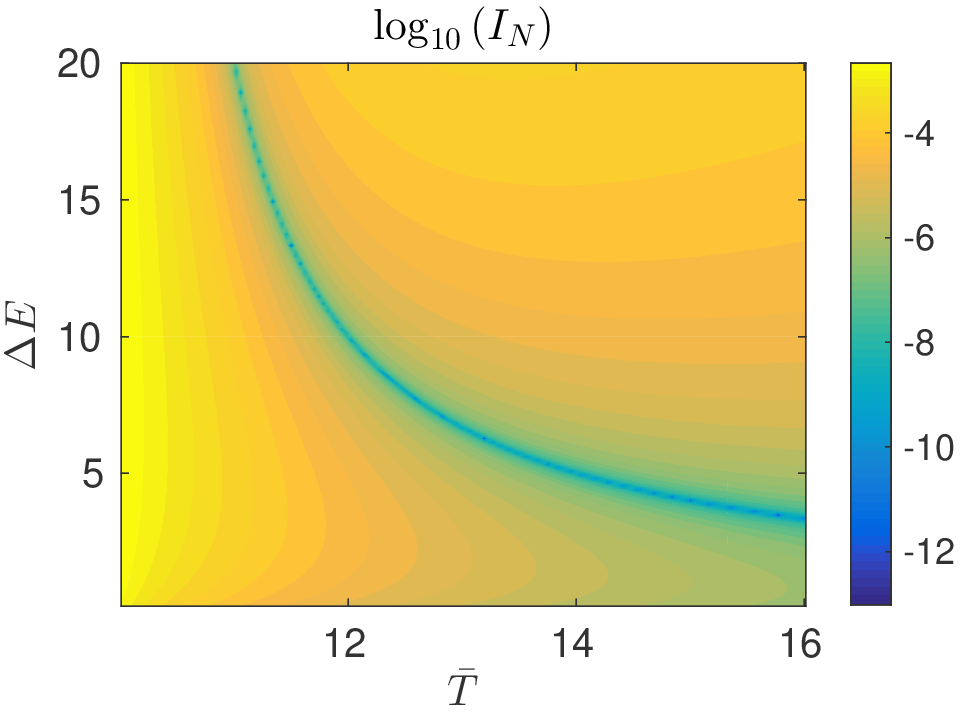}
\caption{  \textit{Mutual Information in the macroscopic limit.} 
We show the mutual information $I_N$ between two spins, each belonging to a different collective spin, for fixed systems parameters $\omega_0 = 0.006, \Gamma_0 = 0.001,\, E_1 = 1,\,\Delta T = T_2 - T_1 = 20$. For this range of parameter, the mutual information between the both spins is qualitatively associated with the power output of the system.
}
\label{fig.PowerOutput.and.MutInf}
\end{figure}

We then analyze the mutual information in the reduced state, which captures the total amount of correlations (both classical and quantum) shared between the two spins. For the two spin system it is defined as $I_N = S(\hat \rho_1) + S(\hat \rho_2) - S(\hat \rho_{1,2})$, where $S(...)$ is the Von Neumann entropy and $\rho_{1(2)} = \mathrm{Tr}_{2(1)}(\rho_{1,2})$ is the single spin reduced density matrix. We show our results in Fig.~\ref{fig.PowerOutput.and.MutInf} for the same range of system parameters as those used in Fig.~\ref{fig.PowerOutput.3rdCum}. We see that the mutual information between the spins is \textit{qualitatively} related to the power output in the system along this region of system parameters, indicating a correspondence between these two quantities. Moreover, we also studied numerically the entanglement between the two spins, as quantified by the concurrence \cite{PhysRevLett.80.2245}. We observe that there is no entanglement, and hence we conclude that the correlations between the two spin do not present any strong quantumness.

\section{Summary and Conclusions}
\label{sec.conclusion}

We have shown collective enhancements of the power output at constant efficiency of a many-body continuous quantum heat engine (with a well-defined macroscopic limit and proper steady-state mode of operation), which can, simultaneously, achieve coherence-enhanced constancy for finite sizes. Such effect appears when ensuring a well-defined macroscopic limit of the model by introducing proper scalings in key engine parameters such as the driving strength, the dissipation rates and the baths termperatures.

In particular, we studied a many-body quantum heat engine 
composed of two ensembles of $N$ spins each, with different energy spacings, collectively dissipating in their respective thermal baths at different temperatures, and subjected to a collective coherent drive able to perform or extract work from the system. The power, the efficiency and the constancy of the collective heat engine has been addressed for the model as a function of the number of spin-pairs $N$, and compared with the case of $N$ separate engines working in parallel.


For the case of no scalings in the engine parameters with $N$, we obtained a super-linear collective gain in the output power, $\mathcal{P}_N/\mathcal{P}_1 \propto N^{\alpha}$, with $\alpha$ in the range $1.0-1.5$.
On the other side, we found that the stability of the system with respect to unavoidable environmental fluctuations (i.e. its constancy) decreases as we consider larger system sizes, hence spoiling the beneficial collective effects. Even in regions where the original two-spins engine overcomes the classical TUR bound, the collective engine tends to loose  stability as we increase $N$, quickly dropping its value below the classical TUR bound.
 
 One of the main merits of our model relies in the fact that it admits a well-defined macroscopic $N \rightarrow \infty$ limit, 
 obtained by performing specific scalings on the engine parameters, and to which we referred 
 as the \textit{high-temperature regime}, see Eq.~\eqref{eq.hight.temp.scaling}. This regime is crucial to explore the persistence of power enhancements for arbitrary large sizes. 
 Within this regime, and by developing a third cumulant approach, we could address the main thermodynamic quantities of the collective engine (power, efficiency and constancy) for both finite sizes and in the macroscopic limit.

 In order to establish the properties of the high-temperature regime, we first analyzed a pure dissipative case, without the coherent driving Hamiltonian ($\omega_0=0$), obtaining 
 analytically the steady state of the system.
 Using a linear stability Jacobian analysis of the dynamical equations, we could also obtain an analytical expression for the the Lindbladian gap in the macroscopic limit. 
 As an interesting remark, we showed that the collective spin system ``feels'' temperature effects only for sufficiently large system sizes $N^*$ which depend algebraically on the bath's temperature, $N^* \sim E/T $. On the contrary, for smaller system sizes the dynamics in the long time limit resembles the one of zero temperature.
 
We then considered the high-temperature regime of the full collective engine model, and analyzed its performance enhancements and constancy. We found that in this case 
there exists a linear gain compared to individual engines working in parallel that survives in the macroscopic limit for certain sets of parameters. Since we  obtained that the efficiency of the engine remains constant 
also in this regime, this implies 
a net gain in the heat engine performance due to the collective nature of their interactions. Our numerical analysis further shows 
that this gain is greater for small average temperatures $\overline{T}$ (with fixed temperature bias $\Delta T$) and large energy splittings ($\Delta E$) between the spins. Interestingly, this region is not directly related to those with a higher power output. 

Quite remarkably, we observed that within the high-temperature regime there are regions in the system parameters where one can have an enhanced stability (and a larger violation of the classical TUR bound) for increasing, but finite, system sizes. In the macroscopic limit, the system local correlations shall resemble classical by the quantum de Finetti theorem. Accordingly, for $N \rightarrow \infty$, we observed a constancy lying always below one, the classical TUR bound. In this context, it would be interesting to perform a  detailed comparison with other scalable versions of quantum heat engines showing TUR violations at the few-body level~\cite{PhysRevB.98.085425,PhysRevE.103.012133,kalaee2021violating}.

Analyzing separately the power output of the system (not specifically its gain), we observed that it is qualitatively related to the mutual information between single spin pairs. 
In the macroscopic limit, the entanglement between pairs of spins in the engine becomes zero, 
pointing again to a classical character of the correlations. 
This analysis may be complemented with other measures of quantum correlations, such as the quantum discord~\cite{Zurek2001}, which can reveal more general forms of quantum correlations among the spins (i.e. while a system may have no entanglement between its constituents, it may still share quantum correlations revealed, for example, by quantum discord quantifier). 
It is worth mentioning that similar models display time-crystalline phases in many-body open systems~\cite{iemini2018,Prazeres2021}. A possible unfolding of our work would be to explore their effects over the operation of the quantum heat engines.

Finally, we remark that, since our model is built by assembling many copies of one of the most fundamental models for quantum thermal machines, we expect our results to be of wide interest in view of future implementations of many-body quantum heat engines. We also expect that our engine model might be amenable of experimental implementation by extending and adapting current implementations of few-body quantum engines using e.g. nuclear spins~\cite{peterson2019}, trapped-ions~\cite{maslennikov2019,lindenfels2019} or cold atoms platforms~\cite{Brantut2013,bouton2021}. In this context, a key point would be to achieve the collective manipulation of the two spin clouds with different linewidths in order to engineer their respective common cold and hot thermal reservoirs, and the interaction of the two ensembles mediated by external driving fields [term $V(t)$ in Eq.~\eqref{eq:V}]. Collective dissipation has been implemented in a number of platforms, including NMR setups~\cite{Viola2001} and cold atoms in optical cavities~\cite{Xu2014,Xu2015,Xu2016}. The later setup is perhaps the most natural candidate where our model could be implemented, following recent proposals for short and long-range dissipation profile engineering~\cite{Seetharam1,Seetharam2} and the possibility of implementing many-body Hamiltonians with long-range interactions~\cite{Henriet2020}. The coherent manipulation of spin ensembles has been also demonstrated in hybrid quantum circuits~\cite{Bianchetti2009,Kubo2011,Nori2013,Okazaki2018,Xu2020} and semiconductor quantum dots~\cite{gangloff2019,gangloff2021}, while similar approaches could be also explored in trapped ions~\cite{Shankar2017,Zhang2020}.

\section{Acknowledgements}
G. M. is founded by Spanish MICINN through the Juan de la Cierva program (IJC2019-039592-I) and acknowledges support from the European Union's Horizon 2020 research and innovation program under the Marie Sk\l{}odowska-Curie grant agreement No 801110 and the Austrian Federal Ministry of Education, Science and Research (BMBWF).
F. I. acknowledges the financial support of the Brazilian funding agencies National Council for Scientific and Technological Development CNPq (Grant No.$308205$/$2019$-$7$) and FAPERJ (Grant No.E-$26$/$211$.$318$/$2019$). Numerical simulations have been performed using the open source QuTiP library~\cite{QuTiP} and MATLAB. 
The corresponding codes have been constructed using well-known tools from these libraries, 
and are available under reasonable request.\\


\appendix
 \section{Derivation of the master equation} \label{sec.master}

 In our model of collective heat engine the environment consist in two independent thermal baths at different temperatures $T_1$ and $T_2$. Bath $1$ is collectively coupled to only the $N$ spin-1/2 systems at energy splitting $E_1$, while bath $2$ is collectively coupled only to the $N$ spin-1/2 systems at $E_2$.  The system-bath interaction reads
 \begin{equation}
 \hat{H}_{SE}= \sum_{i=1}^2 [\hat{S}_{+}^{(i)} \hat{B}_{-}^{(i)} + \hat{S}_{-}^{(i)} \hat{B}_{+}^{(i)}],
 \end{equation}
 where $\hat{B}_{-}^{(i)} = \sum_k \lambda_k^{(i)} \hat{a}^{(i)}_k$ are bath operators, $\hat{B}_{+}^{(i)} = (\hat{B}_{-}^{(i)})^\dagger$, with 
 $[\hat{a}_k^{(i)}, \hat{a}_{k'}^{\dagger (i')}]= \delta_{k, k^\prime} \delta_{i, i^\prime}$ the ladder operators of modes $k,k'$ of the environment and $\lambda_k^{(i)}$ their coupling strengths. Notice that we assume that all spin-1/2 systems are equally coupled to their respective reservoirs. The environmental Hamiltonian is given by 
 \begin{equation}
  \hat{H}_{E} = \sum_{i=1}^2 \sum_{k=1}^\infty \Omega_k^{(i)} \hat{a}_k^{\dagger (i)} \hat{a}_k^{(i)}.
 \end{equation}
 Here for simplicity we assumed bosonic baths. Fermionic bath can be trated similarly by considering ladder operators fulfilling instead $\{\hat{a}_k^{(i)}, \hat{a}_{k'}^{\dagger (i')}\}= \delta_{k, k^\prime}  \delta_{i, i^\prime}$.

 The coupling of the baths to the system spins is assumed to be weak enough such that the linewidth is much smaller than the system energy spacings $E_i$ for $i=1,2$. Each environment is characterized by a spectral density peaked around $E_i$, and is assumed to show an almost Ohmic behavior about the relevant frequency, $J_l(E_i)\simeq \Gamma_0 \delta_{l,i}/2 \pi$, where for simplicity we assume same spontaneous decay rate $\Gamma_0$. Furthermore, we assume a weak driving, where $\omega_0 \sim \sqrt{\Gamma_0}$. Under Born-Markov and secular approximations, a master equation can be then derived using standard techniques in open quantum systems~\cite{breuer2007}, where, due to the above assumptions, the influence of the driving term $\hat{V}_I$ on the dissipators can be neglected. This leads to Eq.~\eqref{eq:masterbtc.pair}, where we obtain two independent dissipators in Lindblad form representing respectively the action of each baths acting collectively on the $N$ spins-1/2 systems with same energy spacing it is coupled to \cite{manzanophd}.

 \section{Collective Thermal Bath in the Macroscopic Limit}
 \label{sec.appendix.collective.bath}
\begin{figure*}
  \includegraphics[width=.45\linewidth]{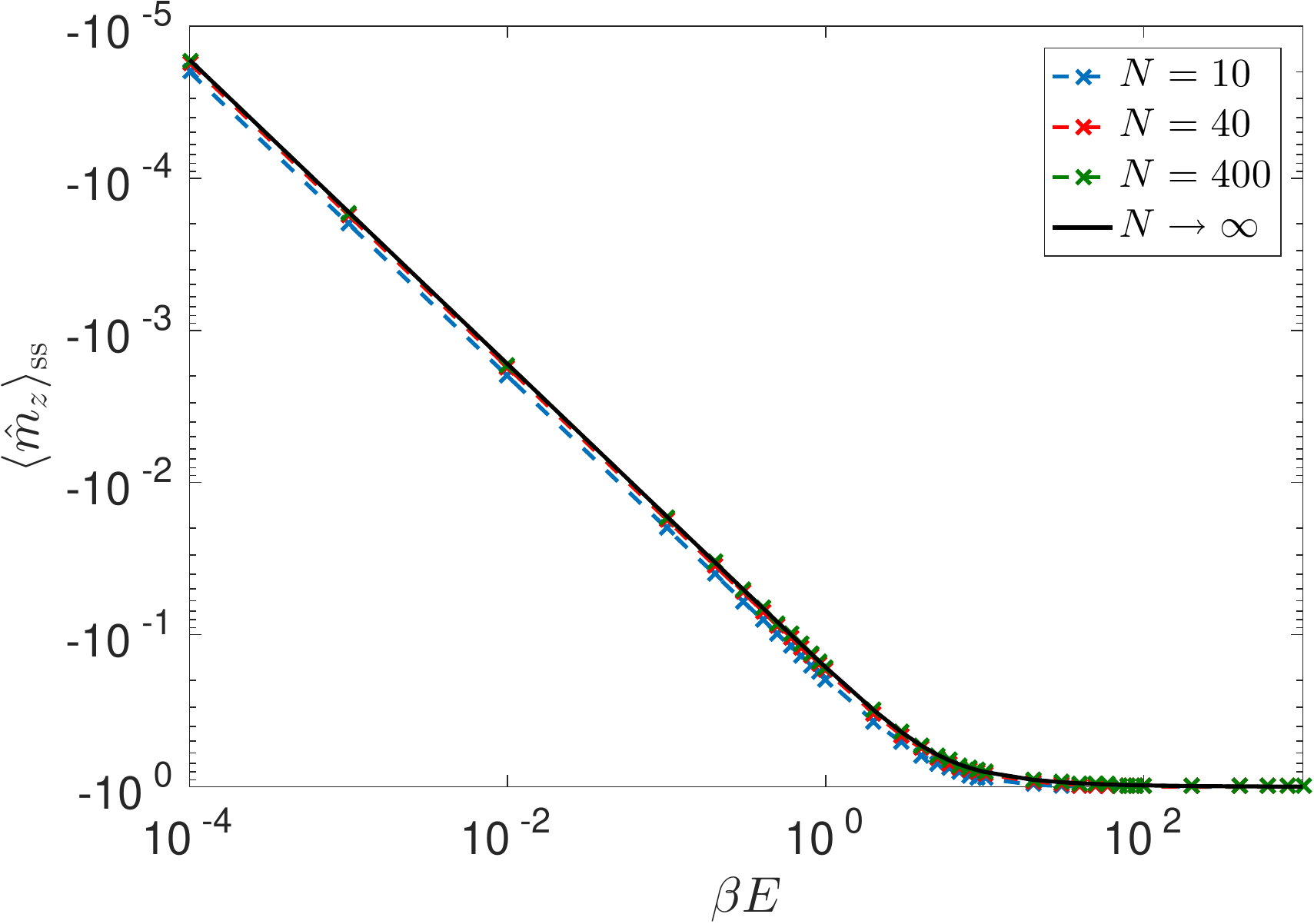}
  \includegraphics[width=.45\linewidth]{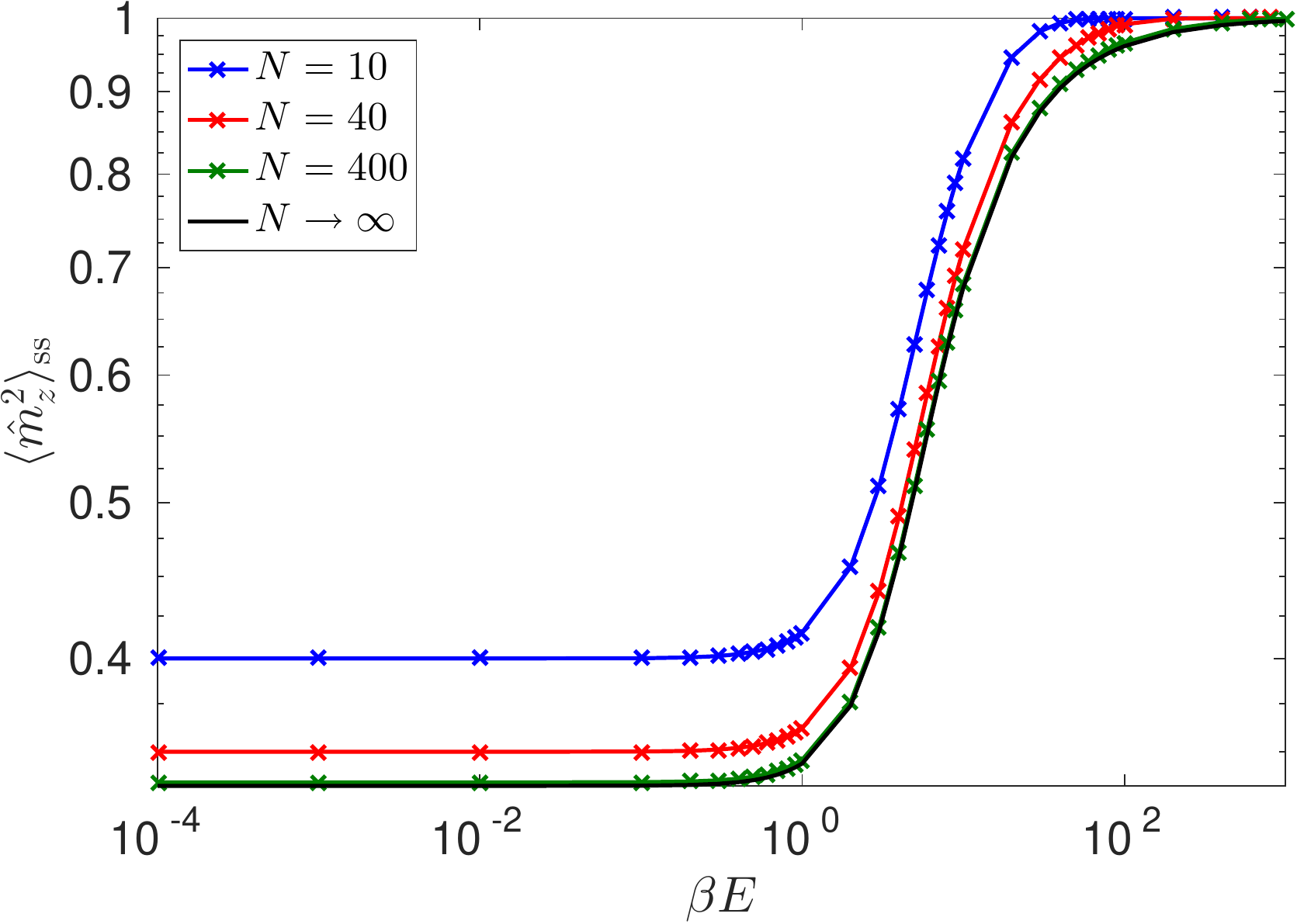}
  \includegraphics[width=.45\linewidth]{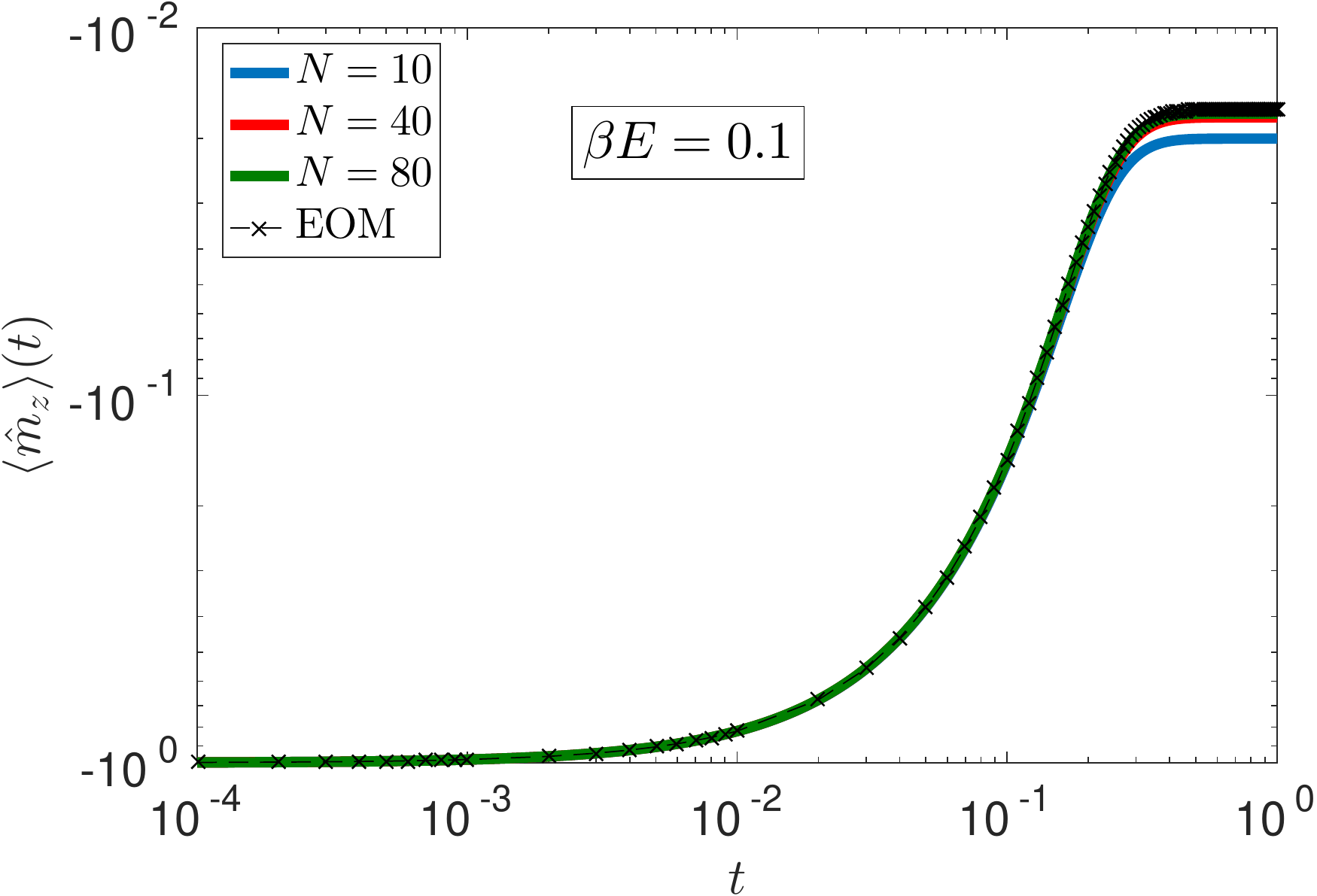}
 \includegraphics[width=.45\linewidth]{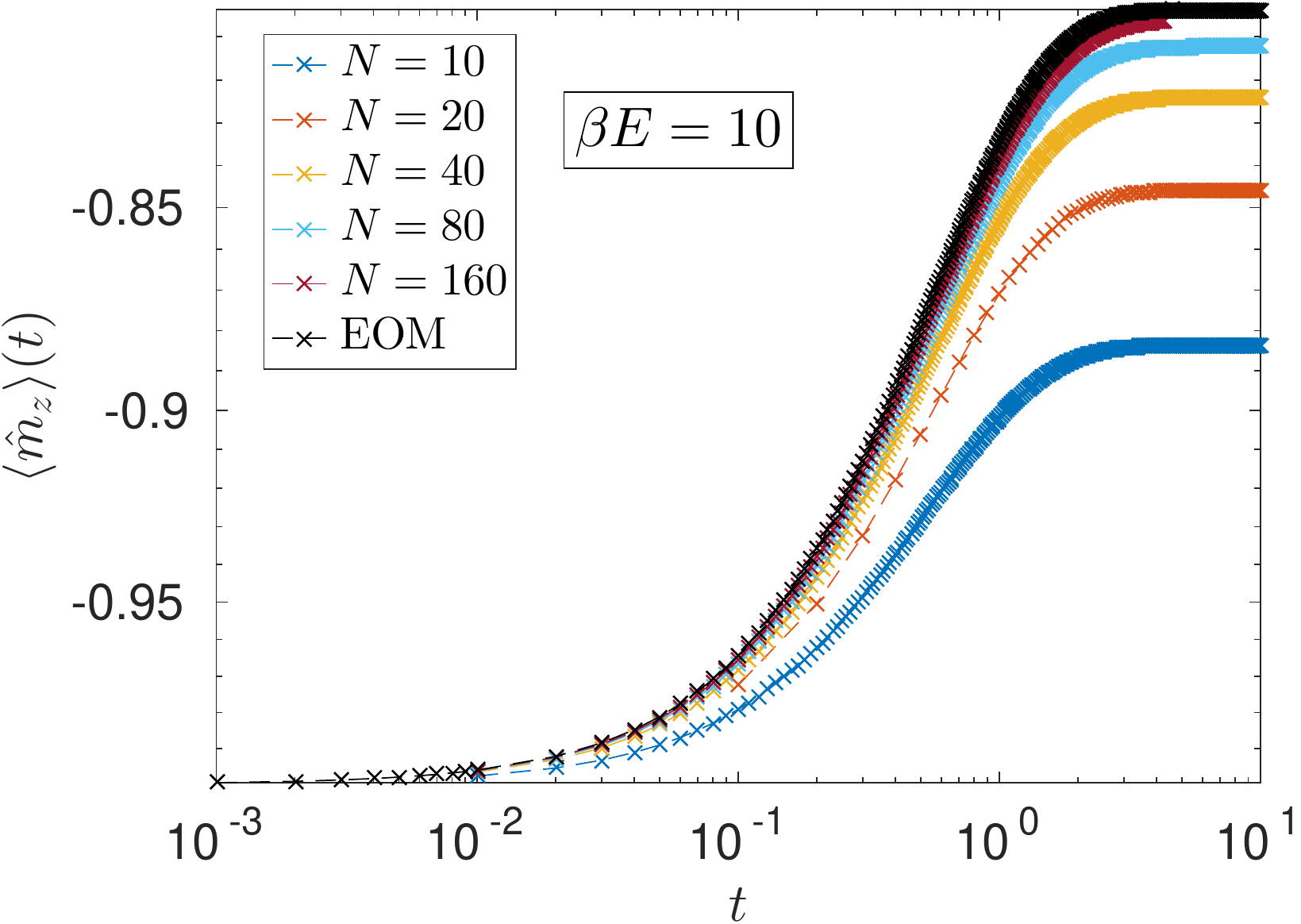}
 \caption{We show in the upper-panels the expectation values for the magnetization $\langle \hat m_z \rangle$ and 
 its quadrature $\langle \hat m_z^2 \rangle$, for different system sizes and temperatures. 
 In the bottom-panels we show the exact dynamics for finite system sizes and those from the effective dynamical equations of motion - Eqs.\eqref{eq:dyn.eq.3rd.dissipation1}-\eqref{eq:dyn.eq.3rd.dissipation2}.  We show the dynamics of the magnetization $\langle \hat m_z \rangle $ for a fixed temperature (bottom-left- panel) $\beta E=0.1$ and (bottom-right-panel) $\beta E=10$. The results show an agreement regarding the macroscopic limit and the finite sizes exact diagonalization trends for the system.
 }
 \label{fig:appendix.collective.thermal.bath}
 \end{figure*}

 We show in this appendix further details on the steady state properties and dynamics for the collective thermal bath of Sec.\eqref{subsec.omega0}.
  We show in Fig.~\ref{fig:appendix.collective.thermal.bath}-top panels the steady state magnetization and variance for finite system sizes as well as in the macroscopic limit. In  
  Fig.~\ref{fig:appendix.collective.thermal.bath}-bottom panels we show the exact dynamics for finite system sizes and those obtained from the effective dynamical equations of motion in the macroscopic limit. We see in both cases an accurate agreement compared to the analytical steady state results and exact diagonalization trends for the dynamics in finite system sizes.

\section{Third Cumulant Equations of Motion for the high temperature regime}
\label{Appendix.Third.Cumulant.pair}

In this Appendix we show the dynamical equations of motions for the two-body observables in the macroscopic limit and high temperature regime (with system parameters scaling as Eq.~\eqref{eq.hight.temp.scaling}). Closing the expectation values at the $3$rd cumulant, we obtain the following effective dynamical equations for the two-body observables:

\begin{widetext}

\begin{eqnarray}\label{3rd.cum.sc.dlxlx}
	\frac{d\langle\hat{m}_{\ell}^{x}\hat{m}_{\ell}^{x}\rangle}{dt}&=&\omega_0\left[\Re\left(\langle \hat{m}_{\ell}^{x}\hat{m}_{\ell}^{z}\rangle\right)\langle\hat{m}_{\bar{\ell}}^{y}\rangle+\left(\langle\hat{m}_{\ell}^{z}\hat{m}_{\bar{\ell}}^{y}\rangle-\langle \hat{m}_{\ell}^{z}\rangle\langle\hat{m}_{\bar{\ell}}^{y}\rangle\right)\langle \hat{m}_{\ell}^{x}\rangle+\left(\langle \hat{m}_{\ell}^{x}\hat{m}_{\bar{\ell}}^{y}\rangle-\langle \hat{m}_{\ell}^{x}\rangle\langle\hat{m}_{\bar{\ell}}^{y}\rangle\right)\langle \hat{m}_{\ell}^{z}\rangle\right]\nonumber\\
	&&+\Gamma_0\left[\langle\left(\hat{m}_{\ell}^{x}\right)^2\rangle\langle\hat{m}_{\ell}^{z}\rangle+2\left(\Re\langle\hat{m}_{\ell}^{x}\hat{m}_{\ell}^{z}\rangle-\langle\hat{m}_{\ell}^{x}\rangle\langle\hat{m}_{\ell}^{z}\rangle\right)\langle\hat{m}_{\ell}^{x}\rangle\right]+\frac{2\Gamma_{0}}{\beta_{\ell}E_{\ell}}\left(\langle\left(\hat{m}_{\ell}^{z}\right)^2\rangle-\langle\left(\hat{m}_{\ell}^{x}\right)^2\rangle\right),
\end{eqnarray}
\begin{eqnarray}\label{3rd.cum.sc.dlxly}
	\frac{d\langle\hat{m}_{\ell}^{x}\hat{m}_{\ell}^{y}\rangle}{dt}&=&\frac{\omega_0}{2}\left[\langle\hat{m}_{\ell}^{z}\hat{m}_{\ell}^{y}\rangle\langle\hat{m}_{\bar{\ell}}^{y}\rangle+\langle \hat{m}_{\ell}^{y}\hat{m}_{\bar{\ell}}^{y}\rangle\langle\hat{m}_{\ell}^{z}\rangle+\langle\hat{m}_{\ell}^{z} \hat{m}_{\bar{\ell}}^{y}\rangle\langle\hat{m}_{\ell}^{y}\rangle-2\langle\hat{m}_{\ell}^{y}\rangle\langle \hat{m}_{\ell}^{z}\rangle\langle\hat{m}_{\bar{\ell}}^{y}\rangle\right.\nonumber\\
	&&\left.\qquad-\left(\langle\hat{m}_{\ell}^{x}\hat{m}_{\ell}^{z}\rangle\langle\hat{m}_{\bar{\ell}}^{x}\rangle+\langle\hat{m}_{\ell}^{x}\hat{m}_{\bar{\ell}}^{x}\rangle\langle\hat{m}_{\ell}^{z}\rangle+\langle\hat{m}_{\ell}^{z}\hat{m}_{\bar{\ell}}^{x}\rangle\langle\hat{m}_{\ell}^{x}\rangle-2\langle\hat{m}_{\ell}^{x}\rangle\langle \hat{m}_{\ell}^{z}\rangle\langle\hat{m}_{\bar{\ell}}^{x}\rangle\right)\right]\nonumber\\
	&&+\frac{\Gamma_0}{2}\left[\left(\langle\hat{m}_{\ell}^{z}\hat{m}_{\ell}^{y}\rangle+\Re\langle\hat{m}_{\ell}^{z}\hat{m}_{\ell}^{y}\rangle\right)\langle\hat{m}_{\ell}^{x}\rangle+\left(\langle\hat{m}_{\ell}^{x}\hat{m}_{\ell}^{z}\rangle+\Re\langle\hat{m}_{\ell}^{x}\hat{m}_{\ell}^{z}\rangle\right)\langle\hat{m}_{\ell}^{y}\rangle\right.\nonumber\\
	&&\left.\qquad+2\Re\left(\langle\hat{m}_{\ell}^{x}\hat{m}_{\ell}^{y}\rangle\right)\langle\hat{m}_{\ell}^{z}\rangle-4\langle\hat{m}_{\ell}^{x}\rangle\langle\hat{m}_{\ell}^{y}\rangle\langle\hat{m}_{\ell}^{z}\rangle\right]-\frac{2\Gamma_{0}}{\beta_{\ell}E_{\ell}}\langle\hat{m}_{\ell}^{x}\hat{m}_{\ell}^{y}\rangle,
\end{eqnarray}
\begin{eqnarray}\label{3rd.cum.sc.dlxlz}
		\frac{d\langle\hat{m}_{\ell}^{x}\hat{m}_{\ell}^{z}\rangle}{dt}&=&\frac{\omega_0}{2}\left[\left(\langle\left(\hat{m}_{\ell}^{z}\right)^2\rangle-\langle\left(\hat{m}_{\ell}^{x}\right)^2\rangle\right)\langle\hat{m}_{\bar{\ell}}^{y}\rangle+2\left(\langle\hat{m}_{\ell}^{z}\hat{m}_{\bar{\ell}}^{y}\rangle-\langle\hat{m}_{\ell}^{z}\rangle\langle\hat{m}_{\bar{\ell}}^{y}\rangle\right)\langle\hat{m}_{\ell}^{z}\rangle-2\left(\langle\hat{m}_{\ell}^{x}\hat{m}_{\bar{\ell}}^{y}\rangle-\langle\hat{m}_{\ell}^{x}\rangle\langle\hat{m}_{\bar{\ell}}^{y}\rangle\right)\langle\hat{m}_{\ell}^{x}\rangle\right.\nonumber\\
		&&\left.\qquad+\langle\hat{m}_{\ell}^{x}\hat{m}_{\ell}^{y}\rangle\langle\hat{m}_{\bar{\ell}}^{x}\rangle+\langle\hat{m}_{\ell}^{x} \hat{m}_{\bar{\ell}}^{x}\rangle\langle\hat{m}_{\ell}^{y}\rangle+\langle\hat{m}_{\ell}^{y} \hat{m}_{\bar{\ell}}^{x}\rangle\langle\hat{m}_{\ell}^{x}\rangle-2\langle\hat{m}_{\ell}^{y}\rangle\langle \hat{m}_{\ell}^{x}\rangle\langle\hat{m}_{\bar{\ell}}^{x}\rangle\right]\nonumber\\
		&&-\frac{\Gamma_0}{2}\left[\left(1-2\langle\left(\hat{m}_{\ell}^{z}\right)^2\rangle\right)\langle\hat{m}_{\ell}^{x}\rangle-4\left(\Re\langle\hat{m}_{\ell}^{x}\hat{m}_{\ell}^{z}\rangle-\langle\hat{m}_{\ell}^{x}\rangle\langle\hat{m}_{\ell}^{z}\rangle\right)\langle\hat{m}_{\ell}^{z}\rangle\right]\nonumber\\
		&&-\frac{\Gamma_{0}}{\beta_{\ell}E_{\ell}}\left(4\Re\langle\hat{m}_{\ell}^{x}\hat{m}_{\ell}^{z}\rangle+\langle\hat{m}_{\ell}^{x}\hat{m}_{\ell}^{z}\rangle\right),
\end{eqnarray}
\begin{eqnarray}\label{3rd.cum.sc.dlyly}
		\frac{d\langle\hat{m}_{\ell}^{y}\hat{m}_{\ell}^{y}\rangle}{dt}&=&-\omega_0\left[\Re\left(\langle \hat{m}_{\ell}^{y}\hat{m}_{\ell}^{z}\rangle\right)\langle\hat{m}_{\bar{\ell}}^{x}\rangle+\left(\langle\hat{m}_{\ell}^{z}\hat{m}_{\bar{\ell}}^{x}\rangle-\langle \hat{m}_{\ell}^{z}\rangle\langle\hat{m}_{\bar{\ell}}^{x}\rangle\right)\langle \hat{m}_{\ell}^{y}\rangle+\left(\langle \hat{m}_{\ell}^{y}\hat{m}_{\bar{\ell}}^{x}\rangle-\langle \hat{m}_{\ell}^{y}\rangle\langle\hat{m}_{\bar{\ell}}^{x}\rangle\right)\langle \hat{m}_{\ell}^{z}\rangle\right]\nonumber\\
		&&+\Gamma_0\left[\langle\left(\hat{m}_{\ell}^{y}\right)^2\rangle\langle\hat{m}_{\ell}^{z}\rangle+2\left(\Re\langle\hat{m}_{\ell}^{y}\hat{m}_{\ell}^{z}\rangle-\langle\hat{m}_{\ell}^{y}\rangle\langle\hat{m}_{\ell}^{z}\rangle\right)\langle\hat{m}_{\ell}^{y}\rangle\right]+\frac{2\Gamma_{0}}{\beta_{\ell}E_{\ell}}\left(\langle\left(\hat{m}_{\ell}^{z}\right)^2\rangle-\langle\left(\hat{m}_{\ell}^{y}\right)^2\rangle\right),
\end{eqnarray}
\begin{eqnarray}\label{3rd.cum.sc.dlylz}
		\frac{d\langle\hat{m}_{\ell}^{y}\hat{m}_{\ell}^{z}\rangle}{dt}&=&-\frac{\omega_0}{2}\left[\left(\langle\left(\hat{m}_{\ell}^{z}\right)^2\rangle-\langle\left(\hat{m}_{\ell}^{y}\right)^2\rangle\right)\langle\hat{m}_{\bar{\ell}}^{x}\rangle+2\left(\langle\hat{m}_{\ell}^{z}\hat{m}_{\bar{\ell}}^{x}\rangle-\langle\hat{m}_{\ell}^{z}\rangle\langle\hat{m}_{\bar{\ell}}^{x}\rangle\right)\langle\hat{m}_{\ell}^{z}\rangle-2\left(\langle\hat{m}_{\ell}^{y}\hat{m}_{\bar{\ell}}^{x}\rangle-\langle\hat{m}_{\ell}^{y}\rangle\langle\hat{m}_{\bar{\ell}}^{x}\rangle\right)\langle\hat{m}_{\ell}^{y}\rangle\right.\nonumber\\
		&&\left.\qquad+\langle\hat{m}_{\ell}^{y}\hat{m}_{\ell}^{x}\rangle\langle\hat{m}_{\bar{\ell}}^{y}\rangle+\langle\hat{m}_{\ell}^{y} \hat{m}_{\bar{\ell}}^{y}\rangle\langle\hat{m}_{\ell}^{x}\rangle+\langle\hat{m}_{\ell}^{x} \hat{m}_{\bar{\ell}}^{y}\rangle\langle\hat{m}_{\ell}^{y}\rangle-2\langle\hat{m}_{\ell}^{x}\rangle\langle \hat{m}_{\ell}^{y}\rangle\langle\hat{m}_{\bar{\ell}}^{y}\rangle\right]\nonumber\\
		&&-\frac{\Gamma_0}{2}\left[\left(1-2\langle\left(\hat{m}_{\ell}^{z}\right)^2\rangle\right)\langle\hat{m}_{\ell}^{y}\rangle-4\left(\Re\langle\hat{m}_{\ell}^{y}\hat{m}_{\ell}^{z}\rangle-\langle\hat{m}_{\ell}^{y}\rangle\langle\hat{m}_{\ell}^{z}\rangle\right)\langle\hat{m}_{\ell}^{z}\rangle\right]\nonumber\\
		&&-\frac{\Gamma_{0}}{\beta_{\ell}E_{\ell}}\left(4\Re\langle\hat{m}_{\ell}^{y}\hat{m}_{\ell}^{z}\rangle+\langle\hat{m}_{\ell}^{y}\hat{m}_{\ell}^{z}\rangle\right),
\end{eqnarray}
\begin{eqnarray}\label{3rd.cum.sc.dlzlz}
		\frac{d\langle\hat{m}_{\ell}^{z}\hat{m}_{\ell}^{z}\rangle}{dt}&=&\omega_0\left[\Re\left(\langle\hat{m}_{\ell}^{y}\hat{m}_{\ell}^{z}\rangle\right)\langle\hat{m}_{\bar{\ell}}^{x}\rangle+\left(\langle \hat{m}_{\ell}^{z}\hat{m}_{\bar{\ell}}^{x}\rangle-\langle \hat{m}_{\ell}^{z}\rangle\langle\hat{m}_{\bar{\ell}}^{x}\rangle\right)\langle\hat{m}_{\ell}^{y}\rangle+\left(\langle \hat{m}_{\ell}^{y}\hat{m}_{\bar{\ell}}^{x}\rangle-\langle \hat{m}_{\ell}^{y}\rangle\langle\hat{m}_{\bar{\ell}}^{x}\rangle\right)\langle\hat{m}_{\ell}^{z}\rangle\right.\nonumber\\
		&&\left.\qquad-\Re\left(\langle \hat{m}_{\ell}^{x}\hat{m}_{\ell}^{z}\rangle\right)\langle\hat{m}_{\bar{\ell}}^{y}\rangle-\left(\langle \hat{m}_{\ell}^{z}\hat{m}_{\bar{\ell}}^{y}\rangle-\langle \hat{m}_{\ell}^{z}\rangle\langle\hat{m}_{\bar{\ell}}^{y}\rangle\right)\langle\hat{m}_{\ell}^{x}\rangle-\left(\langle \hat{m}_{\ell}^{x}\hat{m}_{\bar{\ell}}^{y}\rangle-\langle \hat{m}_{\ell}^{x}\rangle\langle\hat{m}_{\bar{\ell}}^{y}\rangle\right)\langle\hat{m}_{\ell}^{z}\rangle\right]\nonumber\\
		&&-\Gamma_0\left[\left(\langle\left(\hat{m}_{\ell}^{x}\right)^2\rangle+\langle\left(\hat{m}_{\ell}^{y}\right)^2\rangle\right)\langle\hat{m}_{\ell}^{z}\rangle+2\left(\Re\langle\hat{m}_{\ell}^{x}\hat{m}_{\ell}^{z}\rangle-\langle\hat{m}_{\ell}^{x}\rangle\langle\hat{m}_{\ell}^{z}\rangle\right)\langle\hat{m}_{\ell}^{x}\rangle\right.\nonumber\\
		&&\left.\qquad+2\left(\Re\langle\hat{m}_{\ell}^{y}\hat{m}_{\ell}^{z}\rangle-\langle\hat{m}_{\ell}^{y}\rangle\langle\hat{m}_{\ell}^{z}\rangle\right)\langle\hat{m}_{\ell}^{y}\rangle\right]-\frac{2\Gamma_{0}}{\beta_{\ell}E_{\ell}}\left(3\langle\left(\hat{m}_{\ell}^{z}\right)^2\rangle-1\right),
\end{eqnarray}
\begin{eqnarray}\label{3rd.cum.sc.dlxlbarx}
		\frac{d\langle\hat{m}_{\ell}^{x}\hat{m}_{\bar{\ell}}^{x}\rangle}{dt}&=&\frac{\omega_0}{2}\left[\langle \hat{m}_{\ell}^{y}\hat{m}_{\ell}^{x}\rangle\langle\hat{m}_{\bar{\ell}}^{z}\rangle+\langle \hat{m}_{\ell}^{y}\hat{m}_{\bar{\ell}}^{z}\rangle\langle\hat{m}_{\ell}^{x}\rangle+\langle \hat{m}_{\ell}^{x}\hat{m}_{\bar{\ell}}^{z}\rangle\langle\hat{m}_{\ell}^{y}\rangle-2\langle \hat{m}_{\ell}^{y}\rangle\langle\hat{m}_{\ell}^{x}\rangle\langle\hat{m}_{\bar{\ell}}^{z}\rangle\right.\nonumber\\
		&&\left.\qquad+\langle\hat{m}_{\bar{\ell}}^{x}\hat{m}_{\bar{\ell}}^{y}\rangle\langle\hat{m}_{\ell}^{z}\rangle+\langle \hat{m}_{\ell}^{z}\hat{m}_{\bar{\ell}}^{y}\rangle\langle\hat{m}_{\bar{\ell}}^{x}\rangle+\langle\hat{m}_{\ell}^{z}\hat{m}_{\bar{\ell}}^{x}\rangle\langle\hat{m}_{\bar{\ell}}^{y}\rangle-2\langle \hat{m}_{\ell}^{z}\rangle\langle\hat{m}_{\bar{\ell}}^{x}\rangle\langle\hat{m}_{\bar{\ell}}^{y}\rangle\right]\nonumber\\
		&&+\frac{\Gamma_0}{2}\left[\Re\left(\langle\hat{m}_{\ell}^{x}\hat{m}_{\ell}^{z}\rangle\right)\langle\hat{m}_{\bar{\ell}}^{x}\rangle+\langle\hat{m}_{\ell}^{x}\hat{m}_{\bar{\ell}}^{x}\rangle\langle\hat{m}_{\ell}^{z}\rangle+\langle\hat{m}_{\ell}^{z}\hat{m}_{\bar{\ell}}^{x}\rangle\langle\hat{m}_{\ell}^{x}\rangle-2\langle\hat{m}_{\ell}^{x}\rangle\langle\hat{m}_{\ell}^{z}\rangle\langle\hat{m}_{\bar{\ell}}^{x}\rangle\right.\nonumber\\
		&&\left.\qquad+\Re\left(\langle\hat{m}_{\bar{\ell}}^{x}\hat{m}_{\bar{\ell}}^{z}\rangle\right)\langle\hat{m}_{\ell}^{x}\rangle+\langle\hat{m}_{\ell}^{x}\hat{m}_{\bar{\ell}}^{x}\rangle\langle\hat{m}_{\bar{\ell}}^{z}\rangle+\langle\hat{m}_{\ell}^{x}\hat{m}_{\bar{\ell}}^{z}\rangle\langle\hat{m}_{\bar{\ell}}^{x}\rangle-2\langle\hat{m}_{\ell}^{x}\rangle\langle\hat{m}_{\bar{\ell}}^{z}\rangle\langle\hat{m}_{\bar{\ell}}^{x}\rangle\right]\nonumber\\
		&&-\Gamma_{0}\left(\frac{1}{\beta_{\ell}E_{\ell}}+\frac{1}{\beta_{\bar{\ell}}E_{\bar{\ell}}}\right)\langle\hat{m}_{\ell}^{x}\hat{m}_{\bar{\ell}}^{x}\rangle,
\end{eqnarray}
\begin{eqnarray}\label{3rd.cum.sc.dlxlbary}
		\frac{d\langle\hat{m}_{\ell}^{x}\hat{m}_{\bar{\ell}}^{y}\rangle}{dt}&=&\frac{\omega_0}{2}\left[\langle\left(\hat{m}_{\bar{\ell}}^{y}\right)^{2}\rangle\langle\hat{m}_{\ell}^{z}\rangle-\langle\left(\hat{m}_{\ell}^{x}\right)^{2}\rangle\langle\hat{m}_{\bar{\ell}}^{z}\rangle+2\left(\langle\hat{m}_{\ell}^{z}\hat{m}_{\bar{\ell}}^{y}\rangle-\langle\hat{m}_{\ell}^{z}\rangle\langle\hat{m}_{\bar{\ell}}^{y}\rangle\right)\langle\hat{m}_{\bar{\ell}}^{y}\rangle-2\left(\langle\hat{m}_{\ell}^{x}\hat{m}_{\bar{\ell}}^{z}\rangle-\langle\hat{m}_{\ell}^{x}\rangle\langle\hat{m}_{\bar{\ell}}^{z}\rangle\right)\langle\hat{m}_{\ell}^{x}\rangle\right]\nonumber\\
		&&+\frac{\Gamma_0}{2}\left[\Re\left(\langle\hat{m}_{\ell}^{x}\hat{m}_{\ell}^{z}\rangle\right)\langle\hat{m}_{\bar{\ell}}^{y}\rangle+\langle\hat{m}_{\ell}^{x}\hat{m}_{\bar{\ell}}^{y}\rangle\langle\hat{m}_{\ell}^{z}\rangle+\langle\hat{m}_{\ell}^{z}\hat{m}_{\bar{\ell}}^{y}\rangle\langle\hat{m}_{\ell}^{x}\rangle-2\langle\hat{m}_{\ell}^{x}\rangle\langle\hat{m}_{\ell}^{z}\rangle\langle\hat{m}_{\bar{\ell}}^{y}\rangle\right.\nonumber\\
		&&\qquad\left.+\Re\left(\langle\hat{m}_{\bar{\ell}}^{y}\hat{m}_{\bar{\ell}}^{z}\rangle\right)\langle\hat{m}_{\ell}^{x}\rangle+\langle\hat{m}_{\ell}^{x}\hat{m}_{\bar{\ell}}^{y}\rangle\langle\hat{m}_{\bar{\ell}}^{z}\rangle+\langle\hat{m}_{\ell}^{x}\hat{m}_{\bar{\ell}}^{z}\rangle\langle\hat{m}_{\bar{\ell}}^{y}\rangle-2\langle\hat{m}_{\ell}^{x}\rangle\langle\hat{m}_{\bar{\ell}}^{y}\rangle\langle\hat{m}_{\bar{\ell}}^{z}\rangle\right]\nonumber\\
		&&-\Gamma_{0}\left(\frac{1}{\beta_{\ell}E_{\ell}}+\frac{1}{\beta_{\bar{\ell}}E_{\bar{\ell}}}\right)\langle\hat{m}_{\ell}^{x}\hat{m}_{\bar{\ell}}^{y}\rangle,
\end{eqnarray}
\begin{eqnarray}\label{3rd.cum.sc.dlxlbarz}
		\frac{d\langle\hat{m}_{\ell}^{x}\hat{m}_{\bar{\ell}}^{z}\rangle}{dt}&=&\frac{\omega_0}{2}\left[\left(\langle\left(\hat{m}_{\ell}^{x}\right)^{2}\rangle-2\langle\hat{m}_{\ell}^{x}\rangle^2\right)\langle\hat{m}_{\bar{\ell}}^{y}\rangle+\left(\langle\hat{m}_{\bar{\ell}}^{y}\hat{m}_{\bar{\ell}}^{z}\rangle-2\langle\hat{m}_{\bar{\ell}}^{y}\rangle\langle\hat{m}_{\bar{\ell}}^{z}\rangle\right)\langle\hat{m}_{\ell}^{z}\rangle-\left(\langle\hat{m}_{\ell}^{x}\hat{m}_{\ell}^{y}\rangle-2\langle\hat{m}_{\ell}^{x}\rangle\langle\hat{m}_{\ell}^{y}\rangle\right)\langle\hat{m}_{\bar{\ell}}^{x}\rangle\right.\nonumber\\
		&&\qquad\left.+\langle\hat{m}_{\ell}^{z}\hat{m}_{\bar{\ell}}^{y}\rangle\langle\hat{m}_{\bar{\ell}}^{z}\rangle+\langle\hat{m}_{\ell}^{z}\hat{m}_{\bar{\ell}}^{z}\rangle\langle\hat{m}_{\bar{\ell}}^{y}\rangle-\langle\hat{m}_{\ell}^{x}\hat{m}_{\bar{\ell}}^{x}\rangle\langle\hat{m}_{\ell}^{y}\rangle-\langle\hat{m}_{\ell}^{y}\hat{m}_{\bar{\ell}}^{x}\rangle\langle\hat{m}_{\ell}^{x}\rangle+2\langle\hat{m}_{\ell}^{x}\hat{m}_{\bar{\ell}}^{y}\rangle\langle\hat{m}_{\ell}^{x}\rangle\right]\nonumber\\
		&&+\frac{\Gamma_0}{2}\left[\Re\langle\hat{m}_{\ell}^{x}\hat{m}_{\ell}^{z}\rangle\langle\hat{m}_{\bar{\ell}}^{z}\rangle+\langle\hat{m}_{\ell}^{x}\hat{m}_{\bar{\ell}}^{z}\rangle\langle\hat{m}_{\ell}^{z}\rangle+\langle\hat{m}_{\ell}^{z}\hat{m}_{\bar{\ell}}^{z}\rangle\langle\hat{m}_{\ell}^{x}\rangle-2\langle\hat{m}_{\ell}^{x}\rangle\langle\hat{m}_{\ell}^{z}\rangle\langle\hat{m}_{\bar{\ell}}^{z}\rangle\right.\nonumber\\
		&&\qquad\left.-\left(\langle\left(\hat{m}_{\bar{\ell}}^{x}\right)^2\rangle+\langle\left(\hat{m}_{\bar{\ell}}^{y}\right)^{2}\rangle\right)\langle\hat{m}_{\ell}^{x}\rangle-2\left(\langle\hat{m}_{\ell}^{x}\hat{m}_{\bar{\ell}}^{x}\rangle-\langle\hat{m}_{\ell}^{x}\rangle\langle\hat{m}_{\bar{\ell}}^{x}\rangle\right)\langle\hat{m}_{\bar{\ell}}^{x}\rangle-2\left(\langle\hat{m}_{\ell}^{x}\hat{m}_{\bar{\ell}}^{y}\rangle-\langle\hat{m}_{\ell}^{x}\rangle\langle\hat{m}_{\bar{\ell}}^{y}\rangle\right)\langle\hat{m}_{\bar{\ell}}^{y}\rangle\right]\nonumber\\
		&&-\Gamma_{0}\left(\frac{1}{\beta_{\ell}E_{\ell}}+\frac{2}{\beta_{\bar{\ell}}E_{\bar{\ell}}}\right)\langle\hat{m}_{\ell}^{x}\hat{m}_{\bar{\ell}}^{z}\rangle,
\end{eqnarray}
\begin{eqnarray}\label{3rd.cum.sc.dlylbary}
		\frac{d\langle\hat{m}_{\ell}^{y}\hat{m}_{\bar{\ell}}^{y}\rangle}{dt}&=&-\frac{\omega_0}{2}\left[\langle \hat{m}_{\ell}^{x}\hat{m}_{\ell}^{y}\rangle\langle\hat{m}_{\bar{\ell}}^{z}\rangle+\langle \hat{m}_{\ell}^{x}\hat{m}_{\bar{\ell}}^{z}\rangle\langle\hat{m}_{\ell}^{y}\rangle+ \langle\hat{m}_{\ell}^{y}\hat{m}_{\bar{\ell}}^{z}\rangle\langle\hat{m}_{\ell}^{x}\rangle-2\langle\hat{m}_{\ell}^{x}\rangle\langle \hat{m}_{\ell}^{y}\rangle\langle\hat{m}_{\bar{\ell}}^{z}\rangle\right.\nonumber\\
		&&\qquad\left.+\langle\hat{m}_{\bar{\ell}}^{y}\hat{m}_{\bar{\ell}}^{x}\rangle\langle\hat{m}_{\ell}^{z}\rangle+\langle \hat{m}_{\ell}^{z}\hat{m}_{\bar{\ell}}^{x}\rangle\langle\hat{m}_{\bar{\ell}}^{y}\rangle+\langle \hat{m}_{\ell}^{z}\hat{m}_{\bar{\ell}}^{y}\rangle\langle\hat{m}_{\bar{\ell}}^{x}\rangle-2\langle \hat{m}_{\ell}^{z}\rangle\langle\hat{m}_{\bar{\ell}}^{x}\rangle\langle\hat{m}_{\bar{\ell}}^{y}\rangle\right]\nonumber\\
		&&+\frac{\Gamma_0}{2}\left[\Re\left(\langle\hat{m}_{\ell}^{y}\hat{m}_{\ell}^{z}\rangle\right)\langle\hat{m}_{\bar{\ell}}^{y}\rangle+\langle\hat{m}_{\ell}^{y}\hat{m}_{\bar{\ell}}^{y}\rangle\langle\hat{m}_{\ell}^{z}\rangle+\langle\hat{m}_{\ell}^{z}\hat{m}_{\bar{\ell}}^{y}\rangle\langle\hat{m}_{\ell}^{y}\rangle-2\langle\hat{m}_{\ell}^{y}\rangle\langle\hat{m}_{\ell}^{z}\rangle\langle\hat{m}_{\bar{\ell}}^{y}\rangle\right.\nonumber\\
		&&\qquad\left.+\Re\left(\langle\hat{m}_{\bar{\ell}}^{y}\hat{m}_{\bar{\ell}}^{z}\rangle\right)\langle\hat{m}_{\ell}^{y}\rangle+\langle\hat{m}_{\ell}^{y}\hat{m}_{\bar{\ell}}^{y}\rangle\langle\hat{m}_{\bar{\ell}}^{z}\rangle+\langle\hat{m}_{\ell}^{y}\hat{m}_{\bar{\ell}}^{z}\rangle\langle\hat{m}_{\bar{\ell}}^{y}\rangle-2\langle\hat{m}_{\ell}^{y}\rangle\langle\hat{m}_{\bar{\ell}}^{y}\rangle\langle\hat{m}_{\bar{\ell}}^{z}\rangle\right]\nonumber\\
		&&-\Gamma_{0}\left(\frac{1}{\beta_{\ell}E_{\ell}}+\frac{1}{\beta_{\bar{\ell}}E_{\bar{\ell}}}\right)\langle\hat{m}_{\ell}^{y}\hat{m}_{\bar{\ell}}^{y}\rangle,
\end{eqnarray}
\begin{eqnarray}\label{3rd.cum.sc.dlylbarz}
		\frac{d\langle\hat{m}_{\ell}^{y}\hat{m}_{\bar{\ell}}^{z}\rangle}{dt}&=&-\frac{\omega_0}{2}\left[\left(\langle\left(\hat{m}_{\ell}^{y}\right)^{2}\rangle-2\langle\hat{m}_{\ell}^{y}\rangle^2\right)\langle\hat{m}_{\bar{\ell}}^{x}\rangle+\left(\langle\hat{m}_{\bar{\ell}}^{z}\hat{m}_{\bar{\ell}}^{x}\rangle-2\langle\hat{m}_{\bar{\ell}}^{z}\rangle\langle\hat{m}_{\bar{\ell}}^{x}\rangle\right)\langle\hat{m}_{\ell}^{z}\rangle-\left(\langle\hat{m}_{\ell}^{x}\hat{m}_{\ell}^{y}\rangle-2\langle\hat{m}_{\ell}^{x}\rangle\langle\hat{m}_{\ell}^{y}\rangle\right)\langle\hat{m}_{\bar{\ell}}^{y}\rangle\right.\nonumber\\
		&&\qquad\left.+\langle\hat{m}_{\ell}^{z}\hat{m}_{\bar{\ell}}^{x}\rangle\langle\hat{m}_{\bar{\ell}}^{z}\rangle+\langle\hat{m}_{\ell}^{z}\hat{m}_{\bar{\ell}}^{z}\rangle\langle\hat{m}_{\bar{\ell}}^{x}\rangle-\langle\hat{m}_{\ell}^{y}\hat{m}_{\bar{\ell}}^{y}\rangle\langle\hat{m}_{\ell}^{x}\rangle-\langle\hat{m}_{\ell}^{x}\hat{m}_{\bar{\ell}}^{y}\rangle\langle\hat{m}_{\ell}^{y}\rangle+2\langle\hat{m}_{\ell}^{y}\hat{m}_{\bar{\ell}}^{x}\rangle\langle\hat{m}_{\ell}^{y}\rangle\right]\nonumber\\
		&&+\frac{\Gamma_0}{2}\left[\Re\left(\langle\hat{m}_{\ell}^{y}\hat{m}_{\ell}^{z}\rangle\right)\langle\hat{m}_{\bar{\ell}}^{z}\rangle+\langle\hat{m}_{\ell}^{y}\hat{m}_{\bar{\ell}}^{z}\rangle\langle\hat{m}_{\ell}^{z}\rangle+\langle\hat{m}_{\ell}^{z}\hat{m}_{\bar{\ell}}^{z}\rangle\langle\hat{m}_{\ell}^{y}\rangle-2\langle\hat{m}_{\ell}^{y}\rangle\langle\hat{m}_{\ell}^{z}\rangle\langle\hat{m}_{\bar{\ell}}^{z}\rangle\right.\nonumber\\
		&&\qquad\left.-\left(\langle\left(\hat{m}_{\bar{\ell}}^{x}\right)^2\rangle+\langle\left(\hat{m}_{\bar{\ell}}^{y}\right)^{2}\rangle\right)\langle\hat{m}_{\ell}^{y}\rangle-2\left(\langle\hat{m}_{\ell}^{y}\hat{m}_{\bar{\ell}}^{x}\rangle-\langle\hat{m}_{\ell}^{y}\rangle\langle\hat{m}_{\bar{\ell}}^{x}\rangle\right)\langle\hat{m}_{\bar{\ell}}^{x}\rangle-2\left(\langle\hat{m}_{\ell}^{y}\hat{m}_{\bar{\ell}}^{y}\rangle-\langle\hat{m}_{\ell}^{y}\rangle\langle\hat{m}_{\bar{\ell}}^{y}\rangle\right)\langle\hat{m}_{\bar{\ell}}^{y}\rangle\right]\nonumber\\
		&&-\Gamma_{0}\left(\frac{1}{\beta_{\ell}E_{\ell}}+\frac{2}{\beta_{\bar{\ell}}E_{\bar{\ell}}}\right)\langle\hat{m}_{\ell}^{y}\hat{m}_{\bar{\ell}}^{z}\rangle,
\end{eqnarray}
\begin{eqnarray}\label{3rd.cum.sc.dlzlbarz}
		\frac{d\langle\hat{m}_{\ell}^{z}\hat{m}_{\bar{\ell}}^{z}\rangle}{dt}&=&\frac{\omega_0}{2}\left[\left(\langle \hat{m}_{\ell}^{x}\hat{m}_{\ell}^{z}\rangle-\langle \hat{m}_{\ell}^{x}\hat{m}_{\bar{\ell}}^{z}\rangle\right)\langle\hat{m}_{\bar{\ell}}^{y}\rangle+\left(\langle \hat{m}_{\bar{\ell}}^{z}\hat{m}_{\bar{\ell}}^{x}\rangle-\langle \hat{m}_{\ell}^{z}\hat{m}_{\bar{\ell}}^{x}\rangle\right)\langle\hat{m}_{\ell}^{y}\rangle+\left(\langle \hat{m}_{\ell}^{y}\hat{m}_{\bar{\ell}}^{z}\rangle-\langle \hat{m}_{\ell}^{y}\hat{m}_{\ell}^{z}\rangle\right)\langle\hat{m}_{\bar{\ell}}^{x}\rangle\right.\nonumber\\
		&&\qquad\left.+\left(\langle \hat{m}_{\ell}^{z}\hat{m}_{\bar{\ell}}^{y}\rangle-\langle \hat{m}_{\bar{\ell}}^{z}\hat{m}_{\bar{\ell}}^{y}\rangle\right)\langle\hat{m}_{\ell}^{x}\rangle+\left(\langle \hat{m}_{\ell}^{x}\hat{m}_{\bar{\ell}}^{y}\rangle-\langle \hat{m}_{\ell}^{y}\hat{m}_{\bar{\ell}}^{x}\rangle-2\langle \hat{m}_{\ell}^{x}\rangle\langle\hat{m}_{\bar{\ell}}^{y}\rangle+2\langle \hat{m}_{\ell}^{y}\rangle\langle\hat{m}_{\bar{\ell}}^{x}\rangle\right)\left(\langle\hat{m}_{\ell}^{z}\rangle-\langle\hat{m}_{\bar{\ell}}^{z}\rangle\right)\right]\nonumber\\
		&&-\frac{\Gamma_0}{2}\left[\left(\langle\left(\hat{m}_{\ell}^{x}\right)^2\rangle+\langle\left(\hat{m}_{\ell}^{y}\right)^{2}\rangle\right)\langle\hat{m}_{\bar{\ell}}^{z}\rangle+2\left(\langle\hat{m}_{\ell}^{x}\hat{m}_{\bar{\ell}}^{z}\rangle-\langle\hat{m}_{\ell}^{x}\rangle\langle\hat{m}_{\bar{\ell}}^{z}\rangle\right)\langle\hat{m}_{\ell}^{x}\rangle+2\left(\langle\hat{m}_{\ell}^{y}\hat{m}_{\bar{\ell}}^{z}\rangle-\langle\hat{m}_{\ell}^{y}\rangle\langle\hat{m}_{\bar{\ell}}^{z}\rangle\right)\langle\hat{m}_{\ell}^{y}\rangle\right.\nonumber\\
		&&\qquad\left.+\left(\langle\left(\hat{m}_{\bar{\ell}}^{x}\right)^2\rangle+\langle\left(\hat{m}_{\bar{\ell}}^{y}\right)^{2}\rangle\right)\langle\hat{m}_{\ell}^{z}\rangle+2\left(\langle\hat{m}_{\ell}^{z}\hat{m}_{\bar{\ell}}^{x}\rangle-\langle\hat{m}_{\ell}^{z}\rangle\langle\hat{m}_{\bar{\ell}}^{x}\rangle\right)\langle\hat{m}_{\bar{\ell}}^{x}\rangle+2\left(\langle\hat{m}_{\ell}^{z}\hat{m}_{\bar{\ell}}^{y}\rangle-\langle\hat{m}_{\ell}^{z}\rangle\langle\hat{m}_{\bar{\ell}}^{y}\rangle\right)\langle\hat{m}_{\bar{\ell}}^{y}\rangle\right]\nonumber\\
		&&-2\Gamma_{0}\left(\frac{1}{\beta_{\ell}E_{\ell}}+\frac{1}{\beta_{\bar{\ell}}E_{\bar{\ell}}}\right)\langle\hat{m}_{\ell}^{z}\hat{m}_{\bar{\ell}}^{z}\rangle,
\end{eqnarray}
with $\ell,\bar{\ell} \in \{1,2\}$ and $\ell \neq \bar{\ell}$.

\end{widetext}

\section{Power Fluctuations in the Macroscopic Limit}
\label{Appendix.power.fluctuations}

Power fluctuations on the steady state of the system (Eq.~\eqref{eq:power.fluctuation}) are explicitly described in the macroscopic limit and high-temperature regime as follows,
\begin{widetext}
\begin{eqnarray}
	\rm{Var}(\mathcal{P}_N/N)
	&=& \frac{\omega_{0}^2 \left( E_2-E_1 \right)^2}{8} \int_{0}^{\infty} (\langle\hat{m}_{1}^{x}(0) \hat{m}_{2}^{y}(0) \hat{m}_{1}^{x} (\tau) \hat{m}_{2}^{y} (\tau) \rangle_{\rm{ss} }-\langle\hat{m}_{1}^{x}(0)\hat{m}_{2}^{y}(0)\hat{m}_{1}^{y}(\tau)\hat{m}_{2}^{x}(\tau)\rangle_{\rm{ss}} \nonumber \\
	& & \qquad\qquad\qquad\qquad\quad-\langle \hat{m}_{1}^{y}(0) \hat{m}_{2}^{x}(0) \hat{m}_{1}^{x}(\tau) \hat{m}_{2}^{y}(\tau) \rangle_{\rm{ss} }+\langle \hat{m}_{1}^{y}(0)\hat{m}_{2}^{x}(0) \hat{m}_{1}^{y}(\tau) \hat{m}_{2}^{x}(\tau) \rangle_{\rm{ss}}\nonumber \\
	& &\qquad\qquad\qquad\qquad\quad -\langle\hat{m}_{1}^{x} \hat{m}_{2}^{y} - \hat{m}_{1}^{y} \hat{m}_{2}^{x} \rangle_{\rm ss}^2) \rm{d \tau}.
\end{eqnarray}
\end{widetext}
Notice that these fluctuations are only defined with the ratio $N$ in the macroscopic limit, for the reasons discussed in the main text.  In order to compute the fluctuations one must compute the expectation value up to four-body correlations. In order to compute within the $3$rd cumulant approach, we recall that once a cumulant order is closed, all of its higher orders are null as well. Therefore, we use the 
$4$th cumulant closure expression to reduce those four-body correlators to lower orders. Specifically,  closing the $4$th cumulant correlations leads to,
\begin{eqnarray}
	\langle\hat{m}^{\alpha}\hat{m}^{\beta}\hat{m}^{\gamma}\hat{m}^{\delta}\rangle_{\rm ss}&=&\langle\hat{m}^{\alpha}\hat{m}^{\beta}\rangle_{\rm ss}\langle\hat{m}^{\gamma}\hat{m}^{\delta}\rangle_{\rm ss}+\langle\hat{m}^{\alpha}\hat{m}^{\gamma}\rangle_{\rm ss}\langle\hat{m}^{\beta}\hat{m}^{\delta}\rangle_{\rm ss}\nonumber\\
	&+&\langle\hat{m}^{\alpha}\hat{m}^{\delta}\rangle_{\rm ss}\langle\hat{m}^{\beta}\hat{m}^{\gamma}\rangle_{\rm ss}.
\end{eqnarray}  
for $\alpha,\beta, \gamma, \delta  \neq z$, where in the above equation we use implicitly the fact that the steady state expectation value of single-body observables $\langle\hat{m}^{x,y}\rangle_{\rm ss}=0$.
Therefore, we can write the fluctuations in terms of up to $2$-body correlations, as follows:
\begin{widetext}
\begin{eqnarray}
	\rm{Var}(\mathcal{P}_N/N) &=&\frac{\omega_{0}^2\left(E_2-E_1\right)^2}{8}\int_{0}^{\infty}(\langle\hat{m}_{1}^{x}(0)\hat{m}_{1}^{x}(\tau)\rangle_{\rm ss}\langle\hat{m}_{2}^{y}(0)\hat{m}_{2}^{y}(\tau)\rangle_{\rm ss}+\langle\hat{m}_{1}^{x}(0)\hat{m}_{2}^{y}(\tau)\rangle_{\rm ss}\langle\hat{m}_{1}^{x}(\tau)\hat{m}_{2}^{y}(0)\rangle_{\rm ss}\nonumber\\
	&&\qquad\qquad\qquad\qquad\quad-\langle\hat{m}_{1}^{x}(0)\hat{m}_{1}^{y}(\tau)\rangle_{\rm ss}\langle\hat{m}_{2}^{y}(0)\hat{m}_{2}^{x}(\tau)\rangle_{\rm ss}-\langle\hat{m}_{1}^{x}(0)\hat{m}_{2}^{x}(\tau)\rangle_{\rm ss}\langle\hat{m}_{1}^{y}(\tau)\hat{m}_{2}^{y}(0)\rangle_{\rm ss}\nonumber\\
	&&\qquad\qquad\qquad\qquad\quad-\langle\hat{m}_{1}^{y}(0)\hat{m}_{1}^{x}(\tau)\rangle_{\rm ss}\langle\hat{m}_{2}^{x}(0)\hat{m}_{2}^{y}(\tau)\rangle_{\rm ss}-\langle\hat{m}_{1}^{y}(0)\hat{m}_{2}^{y}(\tau)\rangle_{\rm ss}\langle\hat{m}_{1}^{x}(\tau)\hat{m}_{2}^{x}(0)\rangle_{\rm ss}\nonumber\\
	&&\qquad\qquad\qquad\qquad\quad+\langle\hat{m}_{1}^{y}(0)\hat{m}_{1}^{y}(\tau)\rangle_{\rm ss}\langle\hat{m}_{2}^{x}(0)\hat{m}_{2}^{x}(\tau)\rangle_{\rm ss}+\langle\hat{m}_{1}^{y}(0)\hat{m}_{2}^{x}(\tau)\rangle_{\rm ss}\langle\hat{m}_{1}^{y}(\tau)\hat{m}_{2}^{x}(0)\rangle_{\rm ss})\rm{d \tau}.\nonumber \\
\end{eqnarray}
\end{widetext}

In order to compute the two-body time correlations $\langle\hat{m}_{j}^{\alpha}(0)\hat{m}_{k}^{\beta}(\tau)\rangle_{ss}$ we can use quantum regression theorem~\cite{breuer2007}. These will lead to dynamical equations with terms up to $3$-body correlations, which can again be reduced to lower orders within the $3$rd order cumulant closure. 
We obtain the following dynamical equations for the two-body time correlations:

\begin{widetext}

\begin{equation}\label{sc.two_point_d1x}
	\frac{d\langle\hat{m}_{\ell}^{x}(\tau)\hat{m}_{j}^{\alpha}(0)\rangle_{\rm ss}}{d\tau}=\frac{1}{2}\omega_0\langle \hat{m}_{\bar{\ell}}^{y}(\tau)\hat{m}_{j}^{\alpha}(0)\rangle_{\rm ss}\langle \hat{m}_{\ell}^{z}\rangle_{\rm ss}+\frac{\Gamma_0}{2}\langle\hat{m}_{\ell}^{x}(\tau)\hat{m}_{j}^{\alpha}(0)\rangle_{\rm ss}\langle \hat{m}_{\ell}^{z}\rangle_{\rm ss} -\frac{\Gamma_{0}}{\beta_{\ell}E_{\ell}}\langle\hat{m}_{\ell}^{x}(\tau)\hat{m}_{j}^{\alpha}(0)\rangle_{\rm ss},	
\end{equation}
\begin{equation}\label{sc.two_point_d1y}
	\frac{d\langle\hat{m}_{\ell}^{y}(\tau)\hat{m}_{j}^{\alpha}(0)\rangle_{\rm ss}}{d\tau}=-\frac{1}{2}\omega_0\langle \hat{m}_{\bar{\ell}}^{x}(\tau)\hat{m}_{j}^{\alpha}(0)\rangle_{\rm ss}\langle \hat{m}_{\ell}^{z}\rangle_{\rm ss}+\frac{\Gamma_0}{2}\langle\hat{m}_{\ell}^{y}(\tau)\hat{m}_{j}^{\alpha}(0)\rangle_{\rm ss}\langle \hat{m}_{\ell}^{z}\rangle_{\rm ss}-\frac{\Gamma_{0}}{\beta_{\ell}E_{\ell}}\langle\hat{m}_{\ell}^{y}(\tau)\hat{m}_{j}^{\alpha}(0)\rangle_{\rm ss},	
\end{equation}
\end{widetext}
for $\ell,\bar{\ell} \in \{1,2\}$ with $\ell \neq \bar{\ell}$, $j\in \{1,2\}$ and $\alpha = x,y$, where we use $\langle\hat{m}_{j}^{x,y}\rangle_{\rm ss}=0$. These time correlations correspond to a closed set of \textit{linear }dynamical equations (given the input steady state observable 
$\langle m_{j}^{z} \rangle_{\rm ss}$) which can be accurately solved with standard  numerical approaches.

\section{Power enhacements in the macroscopic limit}
\label{appendix.power.enhancement.mac}

\begin{figure}
\includegraphics[width=1 \linewidth]{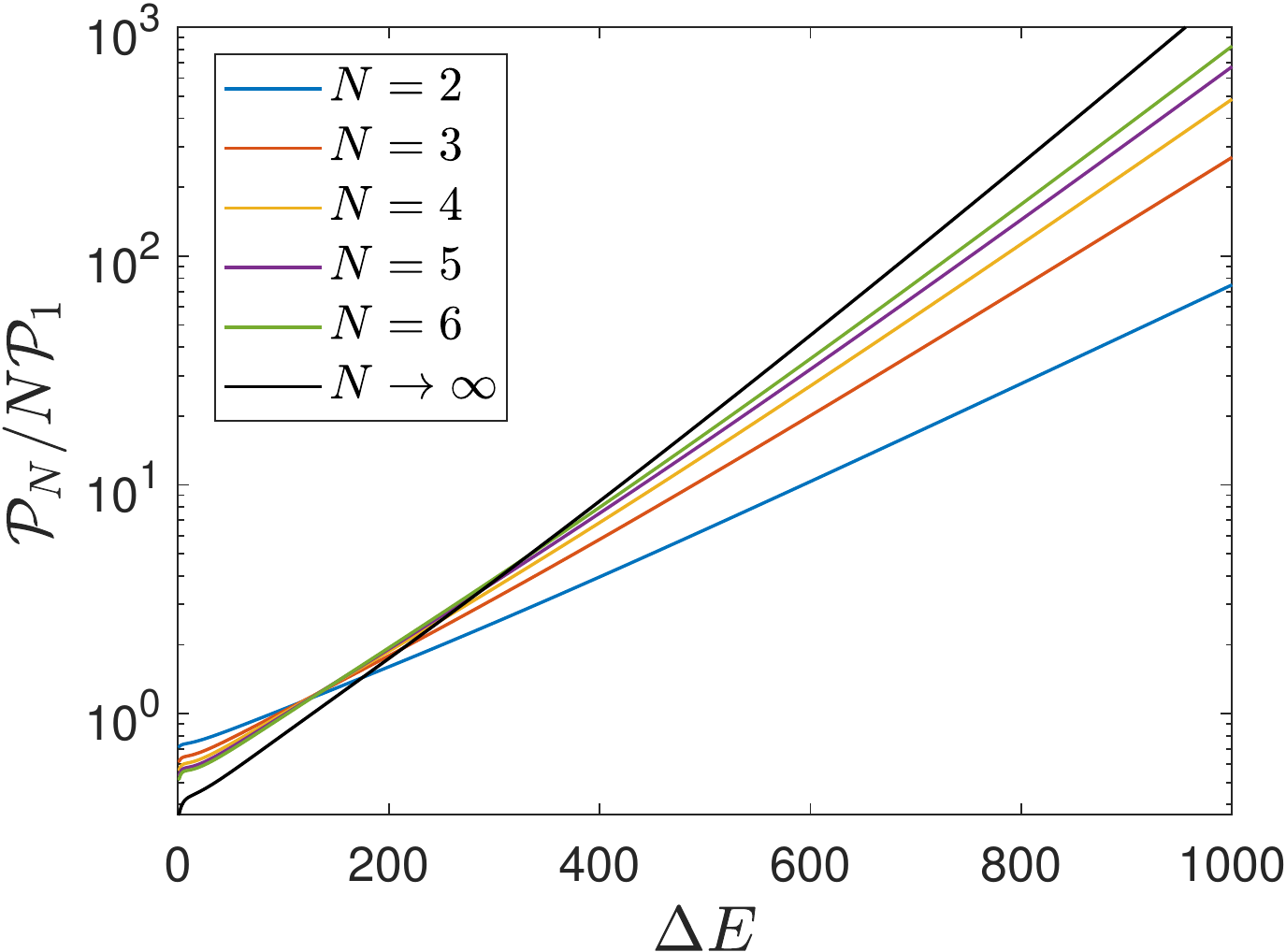}
\caption{  We show the dependence of the power output enhancements with the energy splinting $\Delta E$, for fixed parameters $\omega_0 = 0.006, \Gamma_0 = 0.001,\, E_1 = 1,\, \beta_1=50,\, \beta_2=10^{-2}$, both for finite system sizes and in the macroscopic limit. The collective gain of the power output has an exponential growth with the energy spacing difference.}
\label{fig.appendix.power.ehancement.mac}
\end{figure}

This appendix is dedicated to present some further details concerning the behavior of the power enhancements within the high- temperature regime, complementing the results exposed in Sec.~\eqref{sec.macroscopic.limit}. In particular, in Fig.\eqref{fig.appendix.power.ehancement.mac} we provide evidence that the collective power enhancements $\mathcal{P}_N/(N \mathcal{P}_1)$ dependence with the energy splittings in the system, has an exponential shape with the energy splitting, that is, $\mathcal{P}_N/(N \mathcal{P}_1) \propto e^{\Delta E}$. These results are obtained for both finite sizes and in the macroscopic limit, as can be appreciated from the different lines in Fig.~\ref{fig.appendix.power.ehancement.mac} whose slopes increase with $N$, within the high-temperature regime, as defined by the scalings 
in Eq.~\eqref{eq.hight.temp.scaling}.

\bibliography{QHE-biblio.bib}

\end{document}